\documentclass{article}
\usepackage[utf8]{inputenc}
\usepackage{geometry}
\usepackage[outdir=./]{epstopdf}

\usepackage{auxlib}

\usepackage{makecell}

\usepackage{authblk}

\def\DTAP/{DTAP}

\usepackage{biblatex}
\title{Online Dynamic Acknowledgement with Learned Predictions\thanks{All authors (ordered alphabetically) have equal contributions and are corresponding authors. 
}}

\author[1]{Sungjin Im}
\author[2]{Benjamin Moseley}
\author[3]{Chenyang Xu\thanks{This work was done when the author was a student at Zhejiang University.}}
\author[4]{Ruilong Zhang\thanks{This work was done when the author visited Carnegie Mellon University.}}

\affil[1]{\footnotesize Electrical Engineering and Computer Science, University of California at Merced}
\affil[2]{\footnotesize Tepper School of Business, Carnegie Mellon University}
\affil[3]{\footnotesize Software Engineering Institute, East China Normal University}
\affil[4]{\footnotesize Department of Computer Science, City University of Hong Kong}
\affil[ ]{\texttt{sim3@ucmerced.edu, moseleyb@andrew.cmu.edu, cyxu@sei.ecnu.edu.cn, ruilzhang4-c@my.cityu.edu.hk}}

\date{}

\begin{document}



\maketitle

\begin{abstract}
We revisit the online dynamic acknowledgment problem. In the problem, a sequence of requests arrive over time to be acknowledged, and all outstanding requests can be satisfied simultaneously by one acknowledgement. The goal of the problem is to minimize the total request delay plus acknowledgement cost. This elegant model studies the trade-off between acknowledgement cost and waiting experienced by requests. The problem has been well studied and the tight competitive ratios have been determined. For this well-studied problem, we focus on how to effectively use machine-learned predictions to have better performance.

We develop algorithms that perform arbitrarily close to the optimum with accurate predictions while concurrently having the guarantees arbitrarily close to what the best online algorithms can offer without access to predictions, thereby achieving simultaneous optimum consistency and robustness. This new result is enabled by our novel prediction error measure. No error measure was defined for the problem prior to our work, and natural measures failed due to the challenge that requests with different arrival times have different effects on the objective. We hope our ideas can be used for other online problems with temporal aspects that have been resisting proper error measures.

\end{abstract}

\newpage

\section{Introduction}
\label{sec:intro}

In a typical communication setting where a client receives a sequence of packets from the server, she needs to acknowledge the receipt of the packets to update the server regarding the current communication status. There are two desirable goals in conflict. On the one hand, the server would like to get prompt feedback from the client, which means the client should make more acknowledgements. On the other hand, acknowledging frequently incurs a huge communication cost, and therefore it is desirable to make fewer acknowledgements, which results in prolonged latency in feedback. Thus, there is a fundamental trade-off between making fewer acknowledgements and reducing acknowledgement latency.


The online dynamic acknowledgement problem\footnote{This problem is also called the Dynamic TCP Acknowledgement problem.} (\DTAP/) is an elegant model that was introduced in \cite{DBLP:conf/stoc/DoolyGS98}
to study the above trade-off of fundamental importance. An instance of \DTAP/ is a sequence of $n$ requests (or demands)
arriving online. Succinctly, it can be represented as 
$(p_t)_{t \in [T]}$, where $p_t$ is the number of demands (equivalently packets or requests) that arrive at time  $t\in[T]$ and it is unknown to the algorithm.
The requests must be acknowledged. 
When the client acknowledges (acks for short), all outstanding\footnote{A request is said to be outstanding if it has arrived yet has not been acknowledged (or equivalently satisfied). } requests are simultaneously satisfied. An outstanding request incurs $1/d$ delay cost each time, where $d$ is an input parameter, and each ack costs 1. 
The objective is to minimize the total ack cost plus the delay cost of all requests. 


The \DTAP/ admits a simple 2-competitive\footnote{An online algorithm is said to be $c$-competitive if its objective is at most $c$ times the optimum for all inputs.} greedy algorithm \cite{DBLP:conf/stoc/DoolyGS98} that acks when the outstanding requests have accumulated delay cost equal to 1. Also there is a $\frac{e}{e-1}$-competitive randomized algorithm \cite{DBLP:conf/stoc/Seiden00,DBLP:conf/stoc/KarlinKR01,DBLP:journals/fttcs/BuchbinderN09}. 
The competitive ratios are tight for both deterministic and randomized algorithms. 
The offline version of the problem can be solved optimally via dynamic programming or linear programming as the LP has no integrality gap \cite{DBLP:journals/fttcs/BuchbinderN09}. 

While \DTAP/ is well understood, the traditional study of online algorithms using competitive ratios is often criticized for its pessimistic view of the instances. On the one hand, optimizing the competitive ratio gives robust guarantees against any possible inputs. On the other hand, algorithms that optimize the competitive ratio could be highly tailored toward working well against worst-case instances, sacrificing performance for typical instances that tend not to be adversarial.



The framework of augmenting discrete optimization algorithms with machine learning  \cite{Kraska,lykouris2018competitive} has recently emerged as a powerful framework for algorithm analysis. Such algorithms leverage machine learned parameters to give beyond-worst case performance guarantees while providing robustness guarantees even when given inaccurate parameters from the machine learning. The goal is to develop algorithms that perform extremely well using ML for typical instances and exhibit robustness against exceptional---even adversarial---instances as traditional worst-case algorithms do. 

This model  has been used for various online problems. The predicted parameters can be used to cope with the uncertainty in the input. For example, caching~\cite{Panigrahy,lykouris2018competitive,Rohatgi,DBLP:conf/approx/Wei20}, buy-or-rent~\cite{anand2020customizing,KPS18}, load balancing \cite{LattanziLMV,li2021online},  scheduling \cite{DBLP:conf/spaa/Im0QP21,BamasMRS20,KPS18,mitzenmacher2020scheduling}, secretary problem \cite{Secretary}, metrical task systems \cite{antoniadis2020online}, to name a few. 

ML augmented algorithms typically take an input $I$ and a prediction $P$ on the input. The prediction may be revealed at the beginning or gradually over time as the input is. There is an error function $\eta(I, P)$ defined to measure the quality of the prediction. The prediction is of high quality when $\eta$ is small. 
The algorithm's objective---if it is to be minimized---is commonly bounded by a quantity of the following form:
\begin{equation}
    \label{eqn:consistency}
    \min\{\alpha \opt(I) +  \beta \eta(I, P), \gamma \opt(I)\},
\end{equation}
 where $\opt(I)$ denotes the optimal objective on input $I$. The algorithm is then said to be $\alpha$-\emph{consistent}  and $\gamma$-\emph{robust}. In other words, the algorithm is almost  $\alpha$-competitive when the prediction is very accurate and always at most $\gamma$-competitive simultaneously.

This paper seeks to study the \DTAP/ assuming we have learned predictions on the arriving requests. Specifically, taking access to prediction, $(\hat p_t)_{t \in [T]}$, we would like to achieve consistency and robustness for a certain error measure $\eta$.

\noindent
\paragraph{Simultaneous Optimum Consistency and Robustness.} 
The above parameters, $\alpha, \beta$, and $\gamma$ are correlated, and the guarantees differ depending on the correlation. Ideally, the guarantees should have the following form.

\begin{equation}
    \label{eqn:tradeoff-absolute}
\min \{(1+ \eps) \opt + c \eta, (c^* + \eps) \opt\},
\end{equation}
where $c^*$ is the best competitive ratio that can be achieved without using predictions and $c$ is a constant depending on $\eps$. That is, we would like to achieve two goals simultaneously: on the one hand, we achieve near optimality when the predictions are almost perfect; on the other hand, we simultaneously achieve the best robustness against any inputs regardless of the prediction quality. We say this guarantee is \emph{simultaneous optimum consistency and robustness}. 

Intuitively, this kind of guarantee can be achieved as follows. We have two algorithms, $A$ that closely follows predictions and $B$ that is robust against all inputs. If we know that $\eta$ is large, then we use $B$---otherwise, $A$. However, $\eta$ is a function that depends on the whole input $I$ and the prediction $P$. Therefore, we can only estimate its value before seeing the entire input. Of course, one can design a trivial algorithm that uses $B$ as soon as she notices that the prediction is not perfect. However, the algorithm will then rarely benefit from predictions. Thus, it is critical to define an error measure that grows graciously as the actual instance deviates from the predictions to be able to develop an algorithm that still outperforms $B$ for mild prediction errors. The algorithm crucially relies on the error measure $\eta$.

The primary goal of this paper is to develop an ML-augmented algorithm for the \DTAP/, which seems to resist a reasonable definition of $\eta$. Further, it is one of the most fundamental online problems with temporal aspects where defining a good error measure has been elusive.

\subsection{Critical Need for Prediction Error}

Bamas \emph{et al.} \cite{DBLP:conf/nips/BamasMS20} gave a very elegant framework to smoothly combine various primal-dual algorithms with an arbitrary solution. In particular, for \DTAP/, they assume access to a complete solution as advice and let the primal-dual algorithm mimic the prediction. At a high level, they increase each outstanding request's `potential' and acks when their aggregated potential justifies it. If the algorithm is behind the prediction for a request, it increases its potential more aggressively to catch up with it. The combined algorithm has cost at most $\min\{ \frac{\lambda}{1- e^{-\lambda}} A, \frac{1}{1- e^{-\lambda}} \opt\}$ for any $\lambda \in (0, 1]$, where $A$ is cost of the given solution on the instance. 

Unfortunately, the result has two critical issues. First, it assumes that we are given a complete solution for the input as advice. Thus, if we are only given the number of demands at each time as a prediction, we still have to devise an algorithm. It is possible to use the optimum solution for the predicted instance as a solution for the actual instance, although it is unclear if it is the best way to use the prediction. Another issue is that it does not provide simultaneous optimum consistency and robustness: To achieve near-optimum consistency, one cannot help but make $\frac{\lambda}{1- e^{-\lambda}} \rightarrow 1$, but it will make $\frac{1}{1- e^{-\lambda}} \rightarrow \infty$, resulting  a poor robust guarantee. The error measure serves as a barometer for the prediction's accuracy, and the algorithm cannot change its behavior agilely without it.  


\subsection{Challenges in Defining Prediction Error}

Despite the critical need for an error measure, it is non-trivial to define for \DTAP/---more broadly, problems that involve temporal aspects; see \cref{sec:other}. This is because requests with different arrival times could have different contributions to the objective, and the interaction between the delay cost and the ack cost is subtle. For instance, a few requests arriving later may make the optimal solution switch most of its cost from acks to delay cost and change the solution structure. 

Intuitively, if the error is small, there should be a solution that is simultaneously good for both the actual and predicted instances. It is not difficult to see that naive error measures fail to satisfy it. For example, say we use the $\ell_1$-norm, the aggregate sum of the prediction error at each time, i.e., $\ell_{1}(I,\hatI)=\sum_{t\in[T]}\abs{p_t-\hat{p_t}}$.
But even if the instance is different from the prediction by only one request, the optimum can change a lot depending on whether the request is close to other requests. In short, this is because the $\ell_1$-norm is oblivious to the arrival times. We discuss other natural error measures in \cref{subsubsec:comparison} in detail and why they are unsatisfactory.

\subsection{Our Contributions}
    \label{sec:contribution}

Our contributions are summarized as follows.

\begin{enumerate}
    \item We give \emph{simultaneous optimum consistency and robustness} for \DTAP/ for the first time (\cref{sec:consistency-main} and \cref{subsec:robustness}).
    \item We propose a \emph{novel error measure}, which enables our algorithm and guarantees (\cref{subsubsec:new-eta}).
    \item We show that the predictions are \emph{learnable} with respect to the error measure (\cref{sec:learnable}).  
    \item Our experiments show that our new algorithm beats known learning augmented algorithms and remains on par with the algorithms with the best competitive ratio at the most time (\cref{sec:experiment}).  The experiments demonstrate the theory is predictive of practice. 
\end{enumerate}

Our main theoretical result is the following for our new error $\eta$, which will be shortly described.

\begin{theorem}
    \label{thm:main}
    For any $\eps > 0$, there is a randomized algorithm whose objective is bounded by 
    $\min \{(1+\eps)\opt + O(\frac{1}{\eps^2}) \eta, (\frac{e}{e-1} + \eps)\opt\}$ in expectation. Further, there is a deterministic algorithm whose objective is bounded by 
    $\min \{(1+\eps)\opt + O(\frac{1}{\eps^2}) \eta, (2 + \eps)\opt\}$.
\end{theorem}
In other words, we obtain simultaneous optimum consistency and robustness both deterministically and randomly. We also complement this result by showing that no deterministic algorithms have a cost smaller than $\min\{(1+\lambda-\epsilon)\cdot\opt(I),(1+\frac{\eta}{\lambda})\}$ for any constant $\lambda>0$ and sufficiently small $\epsilon>0$; see \cref{thm:eta3:lowerbound}.

Our new guarantees and algorithm crucially rely on our novel error measure. 
At a high level, we use the optimal objective to define $\eta$, inspired by
\cite{DBLP:conf/spaa/Im0QP21}. Following their idea, we can try to measure the difference between $\opt( (\max\{p_t, \hat p_t\})_{t})$ and $\opt((\min\{p_t, \hat p_t\})_t)$. The former (the latter, resp.) is the optimal objective assuming the number of requests is the maximum (minimum, resp.) of the actual number and the prediction at each time. Although this satisfies the desiderata proposed by them, Monotonicity and Lipschitzness,\footnote{Monotonicity means the error should get smaller if the predictions are more correct; and Lipschitzness means the error should change as much as the objective to successfully distinguish between good and poor predictions.} it fails to capture the temporal aspects of the problem as requests with different arrival times could have a different effect; see \cref{subsubsec:comparison} for more detail. In particular, we show it could mistakenly label some bad predictions as good, making the error unusable in guiding the algorithm's decision. Therefore, we partition the time horizon and aggregate the error measured in each sub-interval. The maximum aggregate error over all partitions is what we adopt.

Under the new error measure, we successfully design a novel algorithm that quickly switches between exploiting the predictions and running robust algorithms. The algorithm is subtle. At a high level, the algorithm first computes a nearly optimal solution that is stable in that adding extra acks cannot significantly reduce the cost. Intuitively this gives us an interval where we can measure the error without worrying too much about the interaction between the delay and ack costs. 
We set a budget we can use until the first time $t_1$ when the nearly optimal solution acks. Until the time $t_1$, at each time we ack if we are still within the budget and the optimal solution on the actual instance would ack right now. Note that we only loosely follow the prediction as the actual instance could be quite different from the prediction. Rather, we use the budget to figure out how much we should tolerate the errors. If we run out of the budget before $t_1$ we switch to a robust algorithm, which can be the 2-competitive greedy algorithm or the $\frac{e}{e - 1}$-competitive randomized algorithm. The algorithm is recursively defined from $t_1$ or from the time it exhausts the budget. 

The prediction $(\hat p_t)_{t \in [T]}$ we use in our paper is natural and provably learnable. We show that the best prediction can be learned from polynomially many samples in $T$ if instances follow a certain unknown distribution.

\subsection{Other Related Work}
    \label{sec:other}
Balancing the communication cost and delay cost has been studied extensively due to its fundamental importance in communication network, such as multicast acknowledgment~\cite{DBLP:conf/wd/DaldoulAM11,DBLP:conf/soda/BritoKV04,DBLP:conf/infocom/BroshS04,DBLP:journals/ton/MokhtarianJ15}; broadcast scheduling~\cite{DBLP:conf/infocom/LuWLP16,DBLP:conf/infocom/SuT97}, etc.
DAP is one of the elegant models that captures the trade-offs between communication cost and delay cost, and therefore it also has been studied in many previous works~\cite{DBLP:journals/wpc/Al-JubariOAH13,DBLP:journals/wpc/ArmaghaniJKR11,DBLP:journals/jcn/ChengL07,DBLP:conf/infocom/OliveiraB05}.

Due to the explosive volume of work in the area of the learning-augmented algorithm, we only discuss the most closely related work. As discussed, \cite{DBLP:conf/nips/BamasMS20} gave a framework that combines an arbitrary solution and a primal-dual algorithm for various problems, such as online set cover and \DTAP/. More generally, \cite{mahdian2012online} showed how to be competitive against two online algorithms simultaneously. In general, such approaches cannot achieve simultaneous optimum consistency and robustness.

We briefly discuss online problems with temporal aspects. To our knowledge, no work prior to ours assumes predictions on jobs or requests' arrival time. Our error measure is inspired in part by the recent work by \cite{DBLP:conf/spaa/Im0QP21} for non-clairvoyant scheduling where the goal is to better minimize average completion time using predictions on job processing times. However, their work assumes all jobs arrive at \emph{the same time}. Its preceding work \cite{purohit2018improving} uses the same prediction model but a different error measure. For average response time, see \cite{DBLP:conf/stoc/AzarLT21}. For various problems involving latency, see \cite{DBLP:conf/focs/AzarT20} and the pointers therein. For connection to inventory management problems, see \cite{buchbinder2008online}.




\section{Preliminaries \& Prediction Error}
\label{sec:model}

\subsection{Notations}

To formally study the \DTAP/, we set up some notations that will be used throughout the paper. We use a set of points in time to denote a solution $\x=\set{x_1,\ldots,x_k}$ where the time points in the set are sorted in increasing order. We say it is feasible to instance $I=(p_t)_{t\in T}$ if (i) $X \subseteq [T]$ and (ii) $\max \x \geq \arg \max_t [p_t > 0]$. 
Let $n=\argmax_{t}[p_t>0]$.
Let $x_0 = 0$ and let $\sF(I,\x)$ be the objective value of the solution $\x$ applied to instance $I$: 
$$
\sF(I,\x)=\abs{\x}+\frac{1}{d}\cdot\sum_{i=1}^{\abs{\x}}\Big( \sum_{t=x_{i-1}+1}^{x_i}p_t(x_i-t) \Big),
$$
where each additional time unit of latency incurs a cost of $\frac{1}{d}$.
For an arbitrary instance $I$, let $\opt(I)$ denote the optimal solution or its objective depending on the context. Similarly, let $\alg(I)$ denote the solution of our algorithm, which will be discussed later, or its objective.
Let $\sD(I,\x)$ and $\num(I,\x)$ be the total delay cost and ack cost of the $\sF(I,\x)$.



In the analysis, we will frequently partition the time horizon $[1,T]$ according to a solution $\x=\set{x_1,\ldots,x_k}$.
The partition $\cP_{\x}=(P_1,\ldots,P_k)$ is called a (time) partition induced by $\x$ if and only if
$P_i=\set{t\mid x_{i-1}<t\leq x_{i}}$ for all $i\in[k]$.
Note that $\cP$ is a partition of the time interval $[1,n]$. 

\subsection{Basic Properties}

Let $I_1=(p_t^1)_{t\in[n]}$ and $I_2=(p_t^2)_{t\in[n]}$ be two arbitrary instances of \DTAP/.
In the following, we define several operations between $I_1,I_2$.
Note that when $I_1$ and $I_2$ stop at a different time point, one could make them stop at the same time by adding some $0$ packet to the short one.
\begin{definition}
Let $I=(p_t)_{t\in[n]}$ be the generated instance by $I_1,I_2$ via the following operations:
\begin{enumerate}
    \item $I:=\abs{I_1\pm I_2}$, where $I=(p_t)_{t\in[n]}$ such that $p_t=\abs{p_t^1\pm p_t^2},\forall t\in[n]$.
    \item $I:=\max\{I_1,I_2\}$, where $I=(p_t)_{t\in[n]}$ such that $p_t=\max\{p_t^1,p_t^2\},\forall t\in[n]$.
    \item $I:=\min\{I_1,I_2\}$, where $I=(p_t)_{t\in[n]}$ such that $p_t=\min\{p_t^1,p_t^2\},\forall t\in[n]$.
\end{enumerate}
\end{definition}

\begin{definition}
Define the relationships between two instances $I_1,I_2$ as follows:
\begin{enumerate}
    \item $I_1\succeq I_2$ if and only if $p_t^1 \geq p_t^2,\forall t\in[n]$.
    \item $I_1 \preceq I_2$ if and only if $p_t^1\leq p_t^2,\forall t\in[n]$.
\end{enumerate}
\end{definition}


For an arbitrary instance $I$ and a feasible solution $\x$, let $\num(I,\x)$, $\sD(I,\x)$ be the number of acks and the total delay costs, respectively.
Note that $\sF(I,\x)=\num(I,\x)+\sD(I,\x)$.

\begin{proposition}
Let $I=(p_t)_{t\in[n]}$ and $I'=(p_t')_{t\in[n]}$ be two arbitrary instances.
Let $\x\subseteq\N_{>0}$ be a feasible solution. 
\begin{enumerate}
    \item $\sF(I,\opt(I))\leq \sF(I,\opt(I'))$.
    \label{pro:per1}
    \item $\num(I,\x)=\num(I',\x)$.
    \label{pro:per2}
    \item $\sD(I,\x)\leq \frac{n}{d}\cdot\sum_{t\in[n]}p_t$.
    \label{pro:per3}
    \item $I\preceq I' \implies \sD(I,\x)\leq \sD(I',\x) $.
    \label{pro:per4}
    \item $\sD(I+I',\x)=\sD(I,\x)+\sD(I',\x)$.
    \label{pro:per5}
    \item $I\preceq I' \implies \opt(I) \leq \opt(I')$.
    \label{pro:per6}
\end{enumerate}
\label{pro:properties}
\end{proposition}

\begin{observation}
Consider a feasible solution $\x=\set{x_1,\ldots,x_k}$ of an instance $I=(p_t)_{t\in[T]}$. Let $\cS_{\x}=(S_1,\ldots,S_k)$ be the partition induced by $\x$ and define subinstance $I_i := (p_t)_{t\in S_i}$.
We have $\sum_{i\in[k]}\opt(I_i) \leq k-1+\opt(I).$
\label{claim:vertical}
\end{observation}

\begin{proof}
Let $t_i$ be the last time point of subinstance $I_i$. We, w.l.o.g., assume that the optimal solution $X^*$ sends an ack at the last time point, implying that $t_k \in X^*$. Construct a new solution $Y := X^* \cup \{t_1,t_2,\ldots,t_{k-1}\}$. Clearly, $\sF(I,Y)\leq k-1+\opt(I)$ because the number of acks increases at most $k-1$ and the total delay cost is non-increasing.

Now partition solution $Y$ into $k$ groups $Y_1,\ldots,Y_k$, where $Y_i$ is the set of the ack time points in the time interval of $I_i$. Since $t_i\in Y$ for any $i\in [k]$, we have $t_i\in Y_i$, and thus, the objective value of solution $Y$ is exactly the sum of its values in $k$ subinstances, i.e., $\sF(I,Y) = \sum_{i=1}^{k} \sF(I_i,Y_i).$
Then due to $\sF(I_i,Y_i)\geq \opt(I_i)$ for any $i\in [k]$, we prove the claim.
\end{proof}

\subsection{Error Measure}

Given an actual instance $I=(p_t)_{t\in[T]}$ of \DTAP/ and its predicted instance $\hat{I}=(\hat{p}_t)_{t\in[\hatT]}$, we would like to define a sound and effective error measure. 
Let $n=\max\{T,\hatT\}$. We can always assume that both the actual instance and the predicted instance have $n$ time points by adding zero package points.
As discussed in \cref{sec:contribution}, we would like to satisfy the Monotonicity and Lipschitzness properties proposed by \cite{DBLP:conf/spaa/Im0QP21}:

\begin{definition}[\cite{DBLP:conf/spaa/Im0QP21}]
The error function $\err$ is monotone if for any $S\subseteq[n]$, 
\begin{align*}
\err\bigg((p_t)_{t\in[n]},(\hat{p}_t)_{t\in[n]}\bigg) \geq \err\bigg((p_t)_{t\in n},(p_t)_{t\in[S]}\cup(\hat{p}_t)_{t\in [n]\setminus S}\bigg),
\end{align*}
while it has Lipschitzness if
\begin{align*}
\abs{\opt((p_t)_{t\in[n]})-\opt((\hat{p}_t)_{t\in[n]})}
\leq \err\bigg((p_t)_{t\in[n]},(\hat{p}_t)_{t\in[n]}\bigg).
\end{align*}
\label{def:error_prop}
\end{definition}



Intuitively, monotonicity ensures that if more request predictions are correct, then the error must decrease. Lipschitzness ensures that the error measure can upper bound the difference between the optimal values of the actual instance and the predicted instance. 

However, a natural extension of the error measure used in \cite{DBLP:conf/spaa/Im0QP21} exhibits a critical weakness---although it satisfies the two properties---that it labels poor predictions as good. This is because the extension fails to capture the requests' arrival time effectively; see \cref{subsubsec:comparison} for the details. To address this challenge, we propose a novel error measure.

\subsubsection{New Error Measure}
\label{subsubsec:new-eta}

Given an instance $I=(p_t)_{t\in[T]}$ and it prediction $\hat{I}=(\hat{p}_t)_{t\in[\hatT]}$, define $\dO(I,\hatI) :=  (\max\{p_t, \hat p_t\})_{t \in [n]}$ and $\dU(I,\hatI) :=(\min\{p_t, \hat p_t\})_{t \in [n]}$ to be the \textit{overpredicted} and \textit{underpredicted} instances respectively. Assuming that we take the max and min at every time step, we can write $\dO(I,\hatI) = \max\{I,\hat{I}\}$ and $\dU(I,\hatI) = \min\{I,\hat{I}\}$.

Let $I\ang{t_1,t_2}$ be the subinstance of $I$ from time $t_1$ to $t_2$, i.e., $(p_t)_{t\in\set{t_1,\ldots,t_2}}$.
We define a partition $\cP$ of the integer set $[T]$ as follows:
$$
\cP:=\set{L_1 = \set{l_0,\ldots,l_1},L_2 = \set{l_1+1,\ldots,l_2},\cdots,}.
$$
The set of consecutive time steps, $L_i$, is called an \textit{interval}.
A partition $\cP$ is called \textit{non-empty} for  instance $I$ if and only if $I\ang{L_i}$ includes at least one request for every $L_i\in\cP$.
Let $\prod(I)$ be the set of all non-empty partitions for  instance $I$.
Note that different partitions may have a  different number of intervals.
We are now ready to define our error measure.

\begin{definition}{(Error Measure)}
Given an instance $I=(p_t)_{t\in[T]}$ and its predicted instance $\hatI=(\hat{p}_t)_{t\in[\hatT]}$, the error measure is defined as follows:
\begin{align*}
\eta(I,\hatI)=\max_{\cP\in\prod(\dU(I,\hatI))} \sum_{L_i\in\cP}
\Big( \opt\left(\dO(I\ang{L_i},\hatI\ang{L_i})\right)
-\opt\left(\dU(I\ang{L_i},\hatI\ang{L_i})\right) \Big)
\end{align*}
\label{def:eta3}
\end{definition}

To understand $\eta$, for a moment, assume that $\cP$ has only one interval. Then, $\eta$ measures how much the optimal objective changes when the number of requests increases from $\min \{p_t, \hat p_t\}$ to $\max \{p_t, \hat p_t\}$ at all times. Although it satisfies Monotonicity and Lipschitzness, it fails to capture how requests arriving at different times affect the ack times---if they change the ack times significantly, intuitively, the prediction is not so good. Thus, we partition the time horizon $[T]$ into intervals and apply the same measure to each interval in $\cP$. Intuitively, if the error is big for some partition, it means $I$ and $\hat I$ have significantly different optimal solution structures. 

When the parameters are clear in the context, we may write $\eta(I,\hatI)$ as $\eta$ for brevity. We claim the following lemma. 


\begin{lemma}
$\eta(I, \hat I)$ can be efficiently computed by dynamic programming and satisfies both monotonicity and Lipschitzness.
\label{lem:eta3:property}
\end{lemma}

\begin{proof}
We first show the Lipschitzness.
By \cref{lem:eta3:auxi-eta:lowerbound}, we have $\tau([1,n],I,\hatI)\leq \eta$.
By \cref{lem:eta3:auxi-eta}, we have $|\opt(I)-\opt(\hatI)|\leq \tau([1,n],I,\hatI)$.
Thus, we have $|\opt(I)-\opt(\hatI)|\leq \eta$.
Now, we show the Monotonicity.
Fix an arbitrary non-empty partition $\cS=\set{S_1,S_2,\ldots}$ and consider an arbitrary subset $S_i$.
Let $f_i$ and $l_i$ be the first and the last time point of $S_i$.
Let $I_i=I\ang{f_i,l_i}$ and $\hatI_i=\ang{f_i,l_i}$.
Then, the value of $\opt(\dO(I_i,\hatI_i))$ will be decreased and the value of $\opt(\dU(I_i,\hatI_i))$ will be increased if we replace some $\hat{p}_t$ with $p_t$, where $t\in\set{f_i,\ldots,l_i}$.
Thus, $\opt(\dO(I_i,\hatI_i))-\opt(\dU(I_i,\hatI_i))$ will be decreased.
\end{proof}



\subsubsection{Auxiliary Prediction Error}
For the sake of analysis, we define another error measure called \textit{Auxiliary Error}.

\begin{definition}{(Auxiliary Error)}
Given $I=(p_t)_{t\in[T]}$, $\hatI=(\hat{p}_t)_{t\in[\hatT]}$, the auxiliary error in time interval $[t_1,t_2]$ is defined as follows:
\begin{align*}
\tau\left( [t_1,t_2],I,\hatI\right)=\; \opt\left( \dO(I\ang{t_1,t_2},\hatI\ang{t_1,t_2}) \right) 
- \opt\left( \dU(I\ang{t_1,t_2},\hatI\ang{t_1,t_2}) \right)
\end{align*}
\label{def:eta3:auxi}
\end{definition}

Note that $\eta(I,\hatI)=\max_{\cP\in\prod(U(I, \hat I))}\sum_{L_i\in \cP}\left( \tau(L_i,I,\hatI) \right)$.
When the parameters are clear from the context, we may write $\tau([1,n],I,\hatI)$ as $\tau$ for simplicity, where $n= \max \{T,\hatT \}$.
%
%
%
The auxiliary error can be used to lower bound $\eta$ and will be useful for the analysis. 
\begin{observation}
For any instance $I$ and its prediction $\hatI$ and for any non-empty partition $\cP=\set{L_1,L_2,\ldots}$ of $\dU(I,\hatI)$, we have
$
\sum_{L_i\in\cP} \left( \tau(L_i,I,\hatI) \right) \leq \eta(I,\hatI)$.
\label{lem:eta3:auxi-eta:lowerbound}
\end{observation}

In the following, we show that the auxiliary error also satisfies both Monotonicity and Lipschitzness.

\begin{lemma}
Given an arbitrary instance $I=(p_t)_{t\in[n]}$ of \DTAP/ and its prediction instance $\hatI=(\hat{p}_t)_{t\in[n]}$, the auxiliary error $\tau([n],I,\hatI)$ satisfies both Monotonicity and Lipschitzness.
\label{lem:eta3:auxi-eta}
\end{lemma}

\begin{proof}
Let $S\subset [n]$ and  $I_S=(p_t)_{t\in S}$, $\hatI_R=(\hat{p}_t)_{t\in[n]\setminus S}$.
We first show the Monotonicity, i.e., $\tau([1,n],I,\hatI)\geq \tau([1,n],I,I_S\cup \hatI_R)$.
Observe that $\dO(I,\hatI)$ can only be decreased and $\dU(I,\hatI)$ can only be increased if we replace some $\hat{p}_t$ with $p_t$.
Therefore, we have $\dO(I,\hatI)\succeq \dO(I,I_S\cup\hatI_R)$ and $\dU(I,\hatI)\preceq\dU(I,I_S\cup\hatI_R)$.
By the property \ref{pro:per6} of \cref{pro:properties}, we have $\opt(\dO(I,\hatI))\geq \opt(\dO(I,I_S\cup\hatI_R))$ and $\opt(\dU(I,\hatI))\leq\opt(\dU(I,I_S\cup\hatI_R))$.
Thus, $\tau([1,n],I,\hatI)$ satisfies the Monotonicity.
Now, we show the Lipschitzness.
We know that  $\dO(I,\hatI)\succeq I,\hatI$ and $\dU(I,\hatI)\preceq I,\hatI$.
Due to the property \ref{pro:per6} of \cref{pro:properties}, we have $\opt(\dO(I,\hatI))\geq \opt(I),\opt(\hatI)$ and $\opt(\dU(I,\hatI))\leq \opt(I),\opt(\hatI)$.
Thus, the Lipschitzness property directly follows.
\end{proof}


\subsubsection{Comparison with Other Prediction Errors}
\label{subsubsec:comparison}

Below we compare our error measure with other natural measures and discuss their shortcomings. 


\paragraph{Absolute Error:}
$\err(I,\hatI)=\abs{\opt(I)-\opt(\hatI)}$.
This definition measures the difference of the optimal objectives of $I$ and $\hatI$. It satisfies Lipschitzness but violates the Monotonicity:
Consider instance $I=(2d,2d,0)$ and its predicted instance $\hatI=(0,2d,2d)$.
Clearly, $\abs{\opt(I)-\opt(\hatI)}=0$, as $\opt(I) = \opt(\hatI) = 2$.
However, if we replace $\hat{p}_1$ with $p_1$ to get a more accurate prediction $(2d,2d,2d)$, $\err(I, \hatI)$ increases to $1$ from $0$.

\paragraph{$\ell_1$-norm Error:}
$\err(I,\hatI)=\sum_{t\in[T]}\abs{p_t-\hat{p}_t}$.
This definition linearly aggregates the difference of the actual number of requests from prediction over all times. This $\ell_1$-norm measure satisfies Monotonicity but violates Lipschitzness:
Consider instance $I=(p_t = \epsilon)_{t\in[T]}$ where $\frac{\eps n(n-1)}{2d} < 1$; thus the optimal solution only acks at the last time.
The prediction is $\hatI=(\hat{p}_t)_{t\in[T]}$ where 
$\hat{p}_t=0$ for all $t\in[T-1]$ and $\hat{p}_T=\epsilon$. Clearly, the optimal solution for $\hat I$ is the same as that for $I$ but it incurs no delay cost for $\hat I$. 
In this case, we have $\abs{\opt(I)-\opt(\hatI)}=\frac{n(n-1)\epsilon}{2d}$. But, 
$\err(I,\hatI) = (n-1) \eps$, so we have 
$\abs{\opt(I)-\opt(\hatI)}\geq \frac{n}{2d} \err(I,\hatI)$. Thus, we need to add $\Theta(n)$ multiplier to the measure to make it usable. 


\paragraph{Auxiliary Error:}
$\err(I,\hatI)=\tau([1,n],I,\hatI)$.  
This error measures how much the optimal objective changes when increasing $\min \{p_t, \hat p_t\}$ to $\max \{p_t, \hat p_t\}$ at all times.
This error measure satisfies both Monotonicity and Lipschitzness 
but is short of capturing the structure of the optimal solution. In particular, it may tag poor predictions as good.
To see this, consider an instance $I$ where a cluster of requests arrive initially until time $t_1$ and one request arrives at a very late time $t_2$. Then, we can set parameters appropriately, so the optimal solution makes only one ack at time $t_2$ and has cost $2 -\epsilon$. Suppose the prediction $\hat I$ is perfect except at time $t_1$, and $\hat p_{t_1}$ is very large. Then, it is easy to see that acking at both times $t_1$ and $t_2$ is optimal and has cost $2$. Thus, the error is at most $\epsilon$, yet $I$ and $\hat I$ have very different optimal solution structures. We can amplify their structural difference by repeating this example over time. 

More specifically, the instance satisfies $p_t\ne 0$ for all $t\in[t^*]\cup\set{n}$, and $p_t=0$ for all the remaining time points.
Specifically, $\frac{p_t}{d}=\frac{1}{t^*(n-t^*-1)}$ for all $t\in[t^*]$.
We assume that $t^*=\frac{1}{100}n$.
Sending at least two acks gives a total cost of at least 2, while if we send an ack only at the time point $n$, the total cost is $1+\frac{2n-t^*-1}{2(n-t^*-1)}=2-\epsilon$.
Thus, the optimal solution will only send an ack at the last time point and $\opt(I)=2-\epsilon$.
Its predicted instance is $\hatI=(\hat{p}_t)_{t\in[n]}$ such that $\hat{p_{t^*}}$ is big enough and $\hat{p}_t=p_t$ for all $t\in[n]\setminus\set{t^*}$.
Clearly, the optimal solution to $\hatI$ will send an ack at the time $t^*$ and $n$, respectively. 
Then, we have $\opt(\hatI)=2+\epsilon$ and $\tau=\opt(\dO(I,\hatI))-\opt(\dU(I,\hatI))=2\epsilon$.
By copying the above instance many times, we see that although $\tau$ is still very small, the structure of the optimal solution to $\hatI$ and $I$ becomes totally different. 
This implies that $\tau$ fails to capture the structure of the optimal solution in the case.
For our error measure $\eta$ defined in \cref{def:eta3}, we can partition $[n]$ into $P_1$ and $P_2$ such that $P_1=\set{1,\cdots,t^*-1}$ and $P_2=[n]\setminus P_1$.
In the case, we have $\tau(P_1,I,\hatI)=0$ and $\tau(P_2,I,\hatI)=1-\epsilon$.
Thus, $\eta\geq 1-\epsilon$, indicating that our new error measure is better to measure the difference between two instances' structures than the previous error.
This implies that our new error measurement can distinguish the current case.


\section{Consistency Bound}
    \label{sec:consistency-main}

In this section, we present an algorithm and prove the consistency. More precisely, we show that for any instance~$I$, the algorithm always returns a solution with the cost at most
$(1+\epsilon)\cdot \opt(I) + O(\frac{1}{\epsilon})\cdot\eta$ for any $\epsilon>0$. When the prediction error $\eta=0$, the competitive ratio is $(1+\epsilon)$. Later in \cref{subsec:robustness}, we show how to refine the algorithm to obtain robustness simultaneously. 
We begin by introducing a crucial definition that is necessary to understand our algorithmic intuition.

\subsection{Stability of Instances}
\label{sec:algorithms}




Let $\y$ be a feasible solution of an instance $I=(p_t)_{t\in[T]}$, i.e., the set of ack time points. Use $\sD(I,\y)$ to denote the total delay cost of solution $Y$ and define $\Delta(I,\y,t) := 
\sD(I,\y) - \sD(I,\y \cup \set{t})$ to be the decrease of the delay cost when making an extra ack at time $t$ in addition to an existing solution~$\y$. 

\begin{definition}\label{def:stable_interval}
For a parameter $\lambda$, let an instance $I=(p_t)_{t\in[T]}$ be a $\lambda$-stable interval if the solution $\x$ which sends only one ack at time $T$ has $\Delta(I,\x,t)\leq 1-\lambda$, $\forall t\in [T]$.
Define an instance's stability factor as the maximum value of $\lambda$ such that it is a $\lambda$-stable interval.
Further, the instance is stable if its stability factor is at least $0$; otherwise, it is unstable.
\end{definition}

Clearly, any instance where the optimal solution $\x$ sends only one ack at time $T$ is at least $0$-stable because for any time $t$, $\Delta(I,\x,t) \leq 1$; otherwise, making an extra ack at time $t$ decreases the total cost. Suppose that the stability factor of the instance is exactly $0$. Namely, there exists at least one time $t$ such that $\Delta(I,\x,t)=1$. We see that increasing $p_t$ slightly will force the optimal solution to make an extra ack. Thus, for a stable interval, the stability factor $\lambda$ measures how much noise can change the structure of the optimal solution on it. 

Defining stability is critical for our algorithm. Intuitively, when the predicted instance is a stable interval, it is convenient to detect a considerable prediction and justify the extra acks. Thus, if we can partition the predicted instance into several stable intervals without incurring too much cost, handling each stable interval independently gives a consistent algorithm.

In the following, \cref{subsec:eta3:one-interval} presents an algorithm to deal with the case that the predicted instance is a stable interval, and then \cref{subsec:eta3:general-case} shows how to do partitioning in the general case and obtain a desirable competitive ratio.

\subsection{Stable Prediction Case}
\label{subsec:eta3:one-interval}



\begin{algorithm}[tb]
\caption{\hspace{-2pt}{\bf .} Predicted-Budget-Based Algorithm}
\label{alg:eta3:one-interval}
\begin{algorithmic}[1]
\REQUIRE Online Instance $I=(p_t)_{t\in[T]}$, prediction $\hatI=(\hat{p}_t)_{t\in[\hatT]}$, and parameter $\lambda >0$
\ENSURE A feasible solution $\s$
\STATE $\s\leftarrow\emptyset$; $j\leftarrow0$; $t\leftarrow1$.
\STATE Compute $\opt(\hatI)$ for the predicted instance.
\STATE {\color{gray}// Phase-1.}
\label{line:one-interval:condition}
\WHILE{$\sF(I\ang{1,t},\s\cup\set{t})<(1+\lambda)\opt(\hatI) $ and the online instance does not end}
%
\IF{$I\ang{j+1,t+1}$ is unstable}
\STATE $\s\leftarrow\s\cup\set{t}$. {\color{gray}// Send an ack at time point $t$.}
\STATE $j\leftarrow t$.
\ENDIF
\STATE $t\leftarrow t+1$.
\ENDWHILE
\STATE {\color{gray}// Phase-2.}
\IF{the online instance does not end}
\STATE Run the traditional $2$-competitive online algorithm for the remaining instance and let $\y$ be the returned solution. Let $\s\leftarrow\s\cup\y$.
\ENDIF
\RETURN S 
\end{algorithmic}
\end{algorithm}

This subsection considers a special case that the predicted instance is a $\lambda$-stable interval, i.e., the optimal solution $\hatX$ only sends one ack, and $\Delta(\hatI,\hatX,t)\leq 1-\lambda$ for all $t\in[\hatT]$. 
The algorithm is stated in~\cref{alg:eta3:one-interval}. It consists of two phases. The first phase is concerned with ``good" prediction cases, while the second phase runs a traditional competitive algorithm to handle inputs that turn out to be far off from the prediction. The pseudo-code uses the 2-competitive algorithm that is deterministic, but it can also be replaced with the $\frac{e}{e-1}$ competitive algorithm that is randomized. 

To decide if the predictions are reliable or not, the algorithm uses $(1+ \lambda) \opt(\hatI)$ as ``budget". 
The algorithm does not switch to the second phase if it has paid within the budget. 
For each time $t$ in the first phase, the algorithm acks if the subinstance $I\ang{j+1,t+1}$ is unstable, where $j$ is the last time an ack was made. 
Using the fact that the prediction makes only one ack at the end of a single interval in a stable instance, we can show that the prediction has a considerable error. Thus, we can charge the cost for making extra acks to the error. Also, the algorithm makes an ack when it is forced to finish the first phase due to running out of budget. 
It is worth noting that the switched time point may be larger than $\hatT$, which is the last time point in the predicted instance. 

\begin{theorem}
Let $\alg(I)$ be the objective value obtained by \cref{alg:eta3:one-interval}.
If the predicted instance is a $\lambda$-stable interval ($\lambda>0$), 
$
\alg(I) \leq (1+\lambda)\cdot \opt(\hatI) + O(\frac{1}{\lambda})\cdot \tau,
$
where $\tau=\tau([1,n],I,\hatI)$ is the auxiliary error (see~\cref{def:eta3:auxi}), which is a lower bound of the prediction error $\eta$.
\label{thm:eta3:ratio:one-interval}
\end{theorem}

We sketch the proof of \cref{thm:eta3:ratio:one-interval}.
Use $\x$ to denote the solution returned by \cref{alg:eta3:one-interval}. 
As stated in \cref{alg:eta3:one-interval}, the algorithm consists of two phases.
Let $e$ be the last time in the first phase; then $e \in \x$ by the algorithm.
Time $e$ partitions $[T]$ into two parts: $S_a=\set{1,\ldots,e}$ and $S_b=\set{e+1,\ldots,T}$.
It is worth to note that $e$ may equal $T$, which makes $S_b=\emptyset$.  
Define $\x_a:=\set{t\in\x\mid t\leq e}$, $\x_b:=\x\setminus\x_a$, $I_a:=(p_i)_{i\in S_a}$ and $I_b:=(p_i)_{i\in S_b}$.
Then, we can split the objective value into two parts: 
$\alg(I)=\sF(I_a,\x_a)+\sF(I_b,\x_b),$
where the first and second terms are the cost incurred for the algorithm in the first and second phases, respectively.
\cref{thm:eta3:ratio:one-interval} can be proved by the following two lemmas. 

\begin{lemma}
$\sF(I_a,\x_a)\leq (1+\lambda)\cdot\opt(\hatI) \leq (1+\lambda)\cdot (\opt(\hatI)+\tau)$.
\label{lem:eta3:one-interval:key1}
\end{lemma}

\begin{proof}
This lemma is easy to prove because the first phase is budget limited, that is, all cost incurred in this phase is at most the budget $(1+\lambda)\opt(\hatI)$. 
Moreover, if the algorithm enters the second phase, the $\sF(I_a,X_a)=(1+\lambda)\opt(\hatI)$ holds by the definition of the algorithm.
Then due to the Lipschitzness of the auxiliary error (\cref{lem:eta3:auxi-eta:lowerbound}), we have $\opt(\hatI)\leq \opt(I)+\tau$, completing the proof.
\end{proof}

\begin{lemma}
$\sF(I_b,\x_b) \leq 2\cdot \opt(I_b)\leq (2+\frac{4}{\lambda})\cdot \tau $.
\label{lem:eta3:one-interval:key}
\end{lemma}



Proving \cref{lem:eta3:one-interval:key} is a bit subtle. 
In the following, we shall apply \cref{claim:vertical} to the analysis.

\begin{proof}[Proof of \cref{lem:eta3:one-interval:key}]

The first inequality uses the fact that the second phase runs the traditional 2-competitive algorithm. So we only need to focus on the second inequality.
We distinguish two cases according to the existence of phase-2. The first case is that the algorithm is still in phase-1 at time $T$, i.e., $I_b=\emptyset$. Then the second inequality is trivial since $\opt(I_b)=0$.
For the second case that $I_b\neq \emptyset$, since $\{I_a, I_b\}$ is a partition of instance $I$, by \cref{claim:vertical},
$\opt(I_b)\leq 1+\opt(I)-\opt(I_a).$

In phase-1, according to the ``if-condition" in \cref{alg:eta3:one-interval}, the subinstance between any two adjacent acks is a stable interval, implying that the algorithm obtains the optimal solution for each subinstance.
Thus, using \cref{claim:vertical} again gives a lower bound of $\opt(I_a)$: $\opt(I_a)\geq \sF(I_a,\x_a)-k_a+1,$
where $k_a$ is the number of acks in phase-1. Since $I_b\neq \emptyset$, the first phase must run out of the predicted budget, which indicates that $\sF(I_a,\x_a) \approx (1+\lambda)\opt(\hatI)$. Without loss of generality, we can assume $\sF(I_a,\x_a) \geq \opt(\hatI)$. Hence, 
$ \opt(I_b)\leq 1+\opt(I)-\opt(\hatI)+k_a-1\leq \tau + k_a,$ 
where the last inequality is due to the Lipschitzness of $\tau$. 

The remaining piece of proof is to bound the value of $k_a$. Up to this point, we have not used the condition that the predicted instance $\hatI$ is a $\lambda$-stable interval. The following analysis uses this condition to prove $k_a \leq 2\tau/\lambda$.

Use $\{I_1,\ldots,I_{k_a},\ldots,I_k\}$ to denote the subinstances induced by solution $X$. 
Let $T_i=[s_i,t_i]$ be the time interval of subinstance $I_i$. 
Recall that $\dO(I,\hatI)$ is the over-predicted instance $(\max\{p_t,\hatp_t\})_{t\in n}$, where $n=\max\{T,\hatT\}$. 
For brevity, denote $\dO(I,\hatI)$ by $I^O=(p_t^O)_{t\in n}$, where $p_t^O=\max\{p_t,\hatp_t\}$.
Let $\hatX$ be the optimal solution of the predicted instance $\hatI$. Since we assume that $\hatI$ is $\lambda$-stable, $\hatX = \{\hatT\}$. 

\begin{claim}
The optimal solution of instance $I^O$ sends at least one ack in interval $T_i$ for any $i\in[k_a]$.
\label{claim:eta3:one-interval:key2}
\end{claim}

\begin{claim}
For any interval partition $\{\hatI_1,\ldots,\hatI_l\}$ of instance $\hatI$, use $Y_i=\{y_i\}$ to denote the solution that only sends an ack at the last time point of $\hatI_i$. We have
 $\sD(\hatI,\hatX) \leq \sum_{i=1}^{l}\sD(\hatI_i,Y_i) + \sum_{i=1}^{l}\Delta(\hatI,\hatX, y_i).   $
\label{claim:eta3:ratio:key}
\end{claim}

For now, we assume that the above two claims are correct and we shall prove them later. 
Use solution $\y=\{y_1,\ldots,y_l\}$ to denote the optimal solution of instance $I^O$.
By \cref{claim:eta3:one-interval:key2}, we know that solution $Y$ sends at least $k_a$ acks, i.e.,
\begin{equation}\label{eq:0}
   l \geq k_a. 
\end{equation}
Thus, we turn to upper bound the value of $l$. Notice that 
\begin{equation}\label{eq:1}
    l= \opt(I^O)- \sD(I^O,\y),
\end{equation}
Use $\{I^O_1,\ldots,I^O_l\}$ to denote the subinstances of instance $I^O$ induced by solution $\y$. Then, $\sD(I^O,\y) = \sum_{i=1}^{l}\sD(I^O_i,\{y_i\}).$
We further define $\hatI_i$ to be the subinstance of $\hatI$ which shares the same time interval of $I^O_i$. Clearly, 
\begin{equation}\label{eq:2}
    \sD(I^O,\y)= \sum_{i=1}^{l}\sD(I^O_i,\{y_i\}) \geq  \sum_{i=1}^{l}\sD(\hatI_i,\{y_i\}).
\end{equation}

By \cref{claim:eta3:ratio:key}, we can connect the above quantity to $\opt(\hatI)$:
\begin{equation}\label{eq:2-5}
    \sum_{i=1}^{l}\sD(\hatI_i,\{y_i\})\geq \sD(\hatI,\hatX) - \sum_{i=1}^{l}\Delta(\hatI,\hatX,y_i). 
\end{equation}
Due to the assumption that $\hatI$ is a $\lambda$-stable interval and the fact that $\Delta(\hatI,\hatX,y_l) = 0$,
\begin{equation}\label{eq:3}
   \sD(\hatI,\hatX) = \opt(\hatI)-1, 
\end{equation}
and
\begin{equation}\label{eq:4}
    \sum_{i=1}^{l}\Delta(\hatI,\hatX,y_i) \leq (1-\lambda)\cdot (l-1).
\end{equation}

Combining the above inequalities, 
we have
\begin{align*}
    l+\sum_{i=1}^{l}\sD(\hatI_i,&\{y_i\}) \leq \opt(I^O) \tag{Eq.~\eqref{eq:1}~\&~\eqref{eq:2}}& \\
    l + \sD(\hatI,\hatX)  &\leq  \opt(I^O)+ \sum_{i=1}^{l}\Delta(\hatI,\hatX,y_i) \tag{Eq.~\eqref{eq:2-5}} \\
    \lambda \cdot l -\lambda &\leq \opt(I^O)-\opt(\hatI) \tag{Eq.~\eqref{eq:3}~\&~\eqref{eq:4}}\\
    k_a &\leq \frac{\tau}{\lambda} +1 \tag{Eq.~\eqref{eq:0} \& Def.~\ref{def:eta3:auxi}}.
\end{align*}

The last piece is to carefully show that $\tau/\lambda$ is always at least $1$, which implies that $k_a\leq 2\tau/\lambda$ and completes the proof. If $k_a \geq 2$, we have $\tau/\lambda \geq k_a-1\geq 1$. Then if $k_a=1$, the algorithm sends only one ack in phase-1. Thus, $I_a$ is a stable interval and $F(I_a,X_a)=\opt(I_a) < \opt(I^O)$. Since the algorithm runs out of the predicted budget $(1+\lambda)\opt(\hatI)$ and enters phase-2, we have $\opt(I^O)\geq (1+\lambda)\cdot \opt(\hatI)$. Again, due to \cref{def:eta3:auxi}, 
$\tau \geq \lambda \cdot \opt(\hatI) \geq \lambda,$
indicating that $\tau/\lambda$ is still at least~$1$.

\end{proof}

\begin{proof}[Proof of \cref{thm:eta3:ratio:one-interval}]
Combining \cref{lem:eta3:one-interval:key1} and \cref{lem:eta3:one-interval:key}, we have $\alg(I) \leq (1+\lambda)\cdot\opt(\hatI) + (2+\frac{4}{\lambda})\cdot \tau$ which directly proves the theorem.
\end{proof}

Now, we prove \cref{claim:eta3:one-interval:key2} and \cref{claim:eta3:ratio:key} in the following.

\begin{proof}
[Proof of \cref{claim:eta3:one-interval:key2}]
For the sake of contradiction, we assume that there exists an $i\in[k]$ such that $\opt(\dO(I,\hatI))$ does not send the ack at any time point in $S_i$.
\cref{alg:eta3:one-interval} sends an ack at the time $s_i$ because $\opt(I\ang{s_{i-1}+1,s_i+1})$ sends an ack at some time point in $S_i$.
For the notation convenience, we define the instance $I\ang{s_{i-1}+1,s_i+1}$ as $I_{s_i}$ and $\dO(I\ang{s_{i-1}+1,s_i+1},\hatI\ang{s_{i-1}+1,s_i+1})$ as $I^O_{s_i}$.
Suppose that $\opt(I_{s_i})$ sends an ack at the time $t$ where $s_{i-1}+1\leq t\leq s_i$.
Thus, we have $\Delta(I_{s_i},\set{s_i+1},t)\geq 1$.
Since $I_{s_i}^O\succeq I_{s_i}$, we have $\Delta(I_{s_i},\set{s_i+1},t)\geq \Delta(I_{s_i}^O,\set{s_i+1},t)\geq 1$.
Thus, there must exist an optimal solution to $\dO(I,\hatI)$ such that the solution also sends an ack at the time $t$.
\end{proof}

The correctness of \cref{claim:eta3:ratio:key} can be directly implied by the following simple observation (\cref{obs:eta3:ratio:key}).
\begin{observation}
Consider an arbitrary instance $I=(p_t)_{t\in[n]}$ of \DTAP/, let $\x$ be a feasible integral solution such that $\x$ only sends one ack at the last time point, i.e., $\x=\set{n}$.
Let $\y=\set{y_1,\ldots,y_k}$ be another feasible integral solution with $|\y|=k$.
Let $\cH_{\y}=(H_1,\ldots,H_k)$ be the solution partition of $\y$.
Let $h_i$ be the first time point in the set $H_i$.
Let $I_i=(p_t)_{t\in H_i}$.
Then, $\sD(I,\x)=\sum_{i\in[k]}\sD(I_i,\set{y_i})+\sum_{i\in[k-1]}\Delta(I\ang{h_i,n},\x,y_i)$.
\label{obs:eta3:ratio:key}
\end{observation}

\subsection{General Algorithm }
\label{subsec:eta3:general-case}

Now we consider the general case. As mentioned above, the basic idea is partitioning the instance into several stable prediction subinstances and dealing with each subinstance separately. The key challenge here is how to partition the instance such that the sum of the subinstances' optimal values is close to the optimal value of the original instance.
We first give a statement in \cref{subsec:gen-alg:stable} to show that it is not ridiculous that such a partition exists.
And then we present the main algorithm in \cref{subsec:gen-alg:main-alg}.

\subsubsection{$\lambda$-Stable Solution}
\label{subsec:gen-alg:stable}

\begin{definition}\label{def:stable_solution}
For an instance $I$, a feasible solution $\y$ with $k$ acks partitions it into $k$ subinstances $\{I_1,I_2,\ldots,I_k\}$. We say solution $\y$ is $\lambda$-stable if any subinstance induced by it is a $\lambda$-stable interval, i.e., for any time $t$ of the whole instance $I$, $\Delta(I,\y,t)\leq 1-\lambda$.
\end{definition}



\begin{algorithm}[htb]
\caption{\hspace{-2pt}{\bf .} Adaptive Predicted-Budget-Based Algorithm}
\label{alg:eta3:general-case}
\begin{algorithmic}[1]
\REQUIRE Online Instance $I=(p_t)_{t\in[T]}$, prediction $\hat{I}=(\hat{p}_t)_{t\in[\hatT]}$ and parameter $\lambda>0$.
\ENSURE A feasible solution $\x$.
\STATE $\x\leftarrow\emptyset$; $t'\leftarrow 0$.
\STATE Let $\y$ be a $\lambda$-stable $\frac{1}{1-\lambda}$-approximation solution of $\hatI$. 
\label{line:general:appro}
\WHILE{$t'<\hatT$ and $\hatI \ang{t',\hatT}$ is not a $\lambda$-stable interval}
\STATE Let $\hatt$ be the minimum time point in $\y$ which is larger than $t'$.
\STATE Run \cref{alg:eta3:one-interval} on the input $\{I\ang{t'+1,T},\hatI\ang{t'+1,\hatt},\lambda\}$ until the first phase ends. 
\STATE Let $\z$ be the returned solution and $t''$ be the termination time.
\IF{$t''<\hatt$}
\STATE Run the traditional 2-competitive algorithm for instance $I\ang{t''+1,\hatt}$ and let $\y$ be the returned solution
\ENDIF

%
\STATE $\x\leftarrow\x\cup\z\cup \y$; $t'\leftarrow \max\{t'',\hatt\}$.
\ENDWHILE
\IF{$t' < \hatT$ and the online instance does not end}
\STATE Run \cref{alg:eta3:one-interval} on the input $\{I\ang{t'+1,T},\hatI\ang{t'+1,\hatT},\lambda\}$ and let $\z$ be the returned solution.
\STATE $\x\leftarrow\x\cup\z$.
\ENDIF
\RETURN $\x$
\end{algorithmic}
\end{algorithm}

\begin{lemma}\label{lem:eta3:approx}
For any instance $I=(p_t)_{t\in[T]}$ of \DTAP/ and any $\lambda\in [0,1)$, there exists a $\lambda$-stable solution which is $(\frac{1}{1-\lambda})$-approximation and can be computed in polynomial time. 
\end{lemma}

To show \cref{lem:eta3:approx}, we give an algorithm (\cref{alg:eta3:approx}) that transforms an arbitrary optimal solution into a $\lambda$-stable solution. 
The intuition is that we start from an optimal solution and always make an extra ack at time $t$ where the $\Delta$ value is larger than $1-\lambda$. Clearly, each extra ack increases the objective value by at most $\lambda$. Thus, the newly incurred cost is at most $\lambda$ times the number of acks in the new solution, implying the total cost of the new solution can be bounded. 

\begin{algorithm}[htb]
\caption{\hspace{-2pt}{\bf .} $\lambda$-stable Solution Constructor.}
\label{alg:eta3:approx}
\begin{algorithmic}[1]
\REQUIRE Arbitrary instance $I=(p_t)_{t\in[n]}$ of \DTAP/ and a constant $\lambda$.
\ENSURE A feasible solution $\y$ such that $\sF(I,\y)\leq \frac{1}{1-\lambda}\cdot \opt(I)$.
\STATE Compute the optimal solution $\x$ to $I$.
\STATE $\y \leftarrow \x$.
\FOR{$t\in[n]$ such that $t\notin\y$} 
\IF{$\Delta(I,\y,t)>1-\lambda$}
\STATE $\y\leftarrow \y\cup\set{t}$.
\ENDIF
\ENDFOR
\RETURN $\y$.
\end{algorithmic}
\end{algorithm}

\begin{proof}
[Proof of \cref{lem:eta3:approx}]
Let $\x$ be the optimal solution to the instance $I$.
We show a simple algorithm that is described in \cref{alg:eta3:approx} returns $(\frac{1}{1-\lambda})$-approximation solution. 
Let $\x$ be the optimal solution to the instance $I$ and $\y$ be the solution returned by \cref{alg:eta3:approx}.
Suppose that \cref{alg:eta3:approx} has $k$ iterations, i.e., $k$ acks are inserted to the solution $\x$ by \cref{alg:eta3:approx}.
Let $\x_i$ be the solution after $(i-1)$-th iteration of \cref{alg:eta3:approx}.
Consider an arbitrary iteration $r$, let $t$ be the time point that is added to $\x$ in the current iteration.
We know that $\Delta(I,\x_r,t)>1-\lambda$ which implies that $\sD(I,\x_r)-\sD(I,\x_r\cup\set{t})>1-\lambda$.
Note that $\num(I,\x_r\cup\set{t})-\num(I,\x_r)=1$.
Thus, $\sF(I,\x_r\cup\set{t})-\sF(I,\x_r)<\lambda$.
Therefore, we have $\sF(I,\y)-\sF(I,\x)\leq k\cdot\lambda$.
Since $k\leq\sF(I,\y)$, we have $\sF(I,\y)\leq \frac{1}{1-\lambda}\cdot\opt(I)$.
\end{proof}

According to \cref{lem:eta3:approx}, we can easily split the predicted instance $\hatI$ into several $\lambda$-stable intervals and the incurred cost is at most $O(\lambda)\opt(\hatI)$. Then due to the Lipschitzness of our prediction error (\cref{lem:eta3:property}), the cost is at most $O(\lambda)(\opt(I)+\eta)$.


\subsubsection{The Main Algorithm}
\label{subsec:gen-alg:main-alg}

We now give the main algorithm which is shown in \cref{alg:eta3:general-case}.
It iteratively treats an interval of $\lambda$-stable $O(\lambda)$-approximation solution as an instance of the stable prediction case and calls \cref{alg:eta3:one-interval}.

From the description, we see that iterations are handled independently. 
In each iteration, the algorithm always starts by trusting the prediction regardless of the states in the previous iterations and enters the next iteration if (\rom{1}) \cref{alg:eta3:one-interval} runs out of the predicted budget of the current iteration; and (\rom{2}) 
 \cref{alg:eta3:one-interval} has processed requests until the last time point of the current predicted interval. 
Since the algorithm adapts in each iteration, we refer to it as an adaptive algorithm. We claim the following theorem.

\begin{theorem}
For an instance $I=(p_t)_{t\in[T]}$ and its prediction $\hatI=(\hat{p}_t)_{t\in[\hatT]}$, let the $\alg(I)$ be the cost of the solution returned by \cref{alg:eta3:general-case} with $\lambda = \Theta(\eps)$, then we have: $\alg(I) \leq (1+\epsilon)\cdot \opt(I) + O(\frac{1}{\epsilon})\cdot\eta $.
Further, the running time is $O(n^2)$, where $n=\max\{T,\hatT\}$.
\label{thm:eta3:ratio}
\end{theorem}

The analysis follows from aggregating the bounds over all the intervals. 
The proof framework is first to show that $\sum_i(\opt(\hatI_i)) \leq (1+O(\lambda))\cdot \opt (\hatI)$ and $\sum_i \tau_i \leq \eta$, where $\hatI_i$ is the $i$-th stable prediction subinstance and $\tau_i$ is its auxiliary error. 
Then, using \cref{thm:eta3:ratio:one-interval} and \cref{lem:eta3:approx} proves the claimed competitive ratio in \cref{thm:eta3:ratio}. 
For the running time, the most time-consuming part is computing the optimal offline solution of the predicted instance, while all other operations can be implemented in the linear time. Since the optimal solution can be computed by a $O(n^2)$ dynamic programming algorithm~\cite{DBLP:conf/stoc/DoolyGS98}, the running time can be proved easily.
Thus, in the proof of \cref{thm:eta3:ratio}, we focus on the proof of the ratio.

\begin{proof}
[Proof of \cref{thm:eta3:ratio}]
Given an arbitrary instance $I=(p_t)_{t\in[n]}$ of \DTAP/ and its predicted instance $\hatI=(\hat{p}_t)_{t\in[m]}$, we assume that \cref{alg:eta3:general-case} has $k$ rounds and the $i$-th round stop at the $t_i$-th time point.
Let $\opt(\hatI_i)$ be the budget of \cref{alg:eta3:one-interval} in the $i$-th round.
Let $\z=\set{t_i\mid i\in[k]}$.
Note that $\z$ is a feasible solution to the actual instance $I$.
Let $\cH_{\z}=(H_1,\ldots,H_k)$ be the solution partition of $\z$.
Note that $\cH_{\z}$ must be a non-empty partition.
Let $I_i=(p_t)_{t\in H_i}$.
By \cref{thm:eta3:ratio:one-interval}, we have $\alg(I_i)\leq (1+\lambda)\cdot \opt(\hatI_i)+(2+\frac{4}{\lambda})\cdot \tau_i$ for all $i\in[k]$, where $\tau_i=\tau(H_i,I,\hatI)$.
Thus, we have $\sum_{i\in[k]}\alg(I_i)\leq (1+\lambda)\cdot\sum_{i\in[k]}\opt(\hatI_i)+(2+\frac{4}{\lambda})\cdot\sum_{i\in[k]}\tau_i$.
Note that $\alg(I)=\sum_{i\in[k]}\alg(I_i)$.
By \cref{lem:eta3:approx}, we have $\sum_{i\in[k]}\opt(\hatI_i)\leq\frac{1}{1-\lambda}\opt(\hatI)$.
Thus, we have $\alg(I)\leq \frac{1+\lambda}{1-\lambda}\cdot\opt(\hatI)+(2+\frac{4}{\lambda})\cdot\sum_{i\in[k]}\tau_i$.
Since $\opt(\hatI)\leq \opt(I)+\tau([1,\max\{n,m\}],I,\hatI)\leq \opt(I)+\eta$, we have
$\alg(I)\leq \frac{1+\lambda}{1-\lambda}\cdot (\opt(I)+\eta)+(2+\frac{4}{\lambda})\sum_{i\in[k]}\tau_i$.
By \cref{lem:eta3:auxi-eta:lowerbound}, we have $\eta\geq\sum_{i\in[k]}\tau_i$.
Therefore, we have $\alg(I)\leq \frac{1+\lambda}{1-\lambda}\cdot\opt(I)+(\frac{1+\lambda}{1-\lambda}+2+\frac{4}{\lambda})\cdot \eta$.
Thus, we have the ratio: $\alg(I)\leq (1+\epsilon)\cdot \opt(I)+O(\frac{1}{\epsilon})\cdot \eta$.
\end{proof}


\subsection{Optimality of Consistency}

We claim the following theorem to show that the dependence of the prediction error $\eta$ in \cref{alg:eta3:general-case} is almost the best possible. 

\begin{theorem}
Given an instance $I$ and its prediction $\hatI$,
the solution of any deterministic algorithm is at least $\min\{(1+\lambda)\cdot \opt(I),\opt(I)+\frac{\eta}{\lambda} \}$, where $\lambda>0$ is a parameter.
\label{thm:eta3:lowerbound}
\end{theorem}

\begin{proof}
Consider the following hard instance.
The predicted instance only has one time point and we assume that the number of packets is $p_1$, where $\frac{p_1}{d}=\epsilon$ and $\epsilon >0$ is an arbitrarily small value.
The actual instance has $p_1$ packets at time $1$, and all the remaining time points have no packets.
The actual instance may have $p_t$ packets at time $t$ depending on the algorithm's decision.
Assume that an arbitrary algorithm $\alg$ sends an ack at the time $t'$.
Let $\sD(p_1)$ be the delay of the packets in time $1$ in $\alg$'s solution.
If $\sD(p_1)\geq \lambda - \epsilon$, then we set $p_t=0$ in the actual instance, i.e., the actual instance only has $p_1$ packets in time $1$.
Clearly, $\alg(I)\geq \lambda+1-\epsilon$ while $\opt(I)=1$.
Thus, $\alg(I)\geq (\lambda+1-\epsilon)\cdot \opt(I)$ in the current case.
If $\sD(p_1)<\lambda - \epsilon$, set $t=t'+1$, i.e., the actual instance has $p_{t'+1}$ packets in the time $t'+1$.
Clearly, $\alg(I)\geq 2+ \sD(p_1)$.
By the definition of $\eta$, we have $\eta=\sD(p_1)+\epsilon$ since $\opt(\dO(I,\hatI))=\opt(I)=1+\sD(p_1)+\epsilon$ and $\opt(\dU(I,\hatI))=\opt(\hatI)=1$.
Thus, $\alg(I)\geq\opt(I)+1$ since $\sD(p_1)=\epsilon$.
Since $\sD(p_1)<\lambda-\epsilon$, we have $\frac{\eta}{\lambda}<1$.
Thus, we have $\alg(I)\geq \opt(I)+\frac{\eta}{\lambda}$, completing the proof.
\end{proof}

\section{Robustness Bound}
\label{subsec:robustness}

This section refines  \cref{alg:eta3:general-case} to obtain robustness bounds in addition to the consistency bound.
Formally, we shall show the following lemma (\cref{lem:eta3:ratio:robustness}) in this section.

\begin{lemma}
Given an arbitrary instance $I=(p_t)_{t\in[n]}$ of \DTAP/, let $\alg(I)$ be the solution returned by \cref{alg:eta3:robustness}. 
We have $\alg(I)\leq\min\{(1+\epsilon)\cdot\opt(I)+O(\frac{1}{\epsilon^2})\cdot\eta,(2+5\epsilon)\cdot\opt(I)\}$.
\label{lem:eta3:ratio:robustness}
\end{lemma}

We will first discuss the high-level idea in \cref{sec:robust:intuition}.
And then, we state the formal proof in \cref{sec:robust:proof}.

\subsection{Intuition}
\label{sec:robust:intuition}

In this subsection, we discuss the high-level ideas.
For each time $t$, define $I_t:=I\ang{1,t}$ and $\eta_t:=\eta (I_t\cup \hatI\ang{t+1,n},\hatI )$. 
Due to the monotonicity of the error, $\eta_t$ increases as $t$ increases. 
An intuitive way to gain robustness is switching to the 2-competitive deterministic (or $e / (e -1)$-competitive randomized) algorithm when $\eta_t$ is found to be large. 
If we know the optimal value $\opt(I)$, we can then make the switch at the first time $t$ we observe the error is large, i.e., $\eta_t> \epsilon \opt(I)$. Due to the monotonicity discussed above, the actual error $\eta$ will only be large. Thus, we can achieve $(1+\epsilon)$-consistency bound along with $(2 +\epsilon)$-robustness bound (or $e/(e-1)+\epsilon$ if randomization is allowed).


However, we can only see $I_t$ at each time $t$ and thus, $\opt(I)$ is unknown. 
The current error $\eta_t$ could look large compared to $\opt(I_t)$, but turn out to be very small compared to $\opt(I)$. 
We address this issue based on the observation that an instance can be partitioned into several subinstances such that the optimal cost of each subinstance is at most $1/{\epsilon}$ while increasing the aggregate optimal cost by at most $(1+\epsilon)$ factor. 
Further, this partition can be done online by checking the optimal solution to the current subinstance at every time point.
Then, for each subinstance, we can argue that if the current error is large enough to shift to the traditional algorithm, it is also large against the optimum for the subinstance. In this process, to obtain a robustness guarantee, we increase the coefficient of $\eta$ from $O(\frac{1}{\epsilon})$ to $O(\frac{1}{\epsilon^2})$ in the consistency bound for a technical reason. 
Combing \cref{thm:eta3:ratio} and the robustness scheme, \cref{thm:main} can be proved.

\subsection{The Formal Proof}
\label{sec:robust:proof}

In this subsection, we formally show the robustness result stated in \cref{lem:eta3:ratio:robustness}.
Let $\alg(I)$ be the solution returned by \cref{alg:eta3:general-case}.
By \cref{thm:eta3:ratio}, we know that $\alg(I)\leq \frac{1+\lambda}{1-\lambda}\cdot\opt(I)+(\frac{1+\lambda}{1-\lambda}+2+\frac{4}{\lambda})\cdot \eta$.
For notational convenience, let $\frac{1+\lambda}{1-\lambda}+2+\frac{4}{\lambda}=\frac{1}{\epsilon_{\lambda}}$.
Following the intuition above, at each time point, we check the current error, when the error is large the algorithm will be switched to the traditional online algorithm; otherwise, we run \cref{alg:eta3:general-case}.
The formal description can be found in \cref{alg:eta3:robustness}.
In order to further simplify the expression of ratio, we assume that the value of the optimal solution is always less than $\frac{1}{\lambda}$. 
Finally, we use \cref{lem:eta3:robustness:approx} to remove this assumption.

\begin{lemma}
Given an arbitrary instance $I=(p_t)_{t\in[n]}$ of \DTAP/, there exists a set of time points $Q\subseteq[n]$, $Q=\set{q_1,\ldots,q_k}$ with $\opt(I\ang{q_{i-1}+1,q_i})\leq\frac{1}{\lambda}$ such that $\sum_{i=1}^{k}\opt(I\ang{q_{i-1}+1,q_i}))\leq (1+\lambda)\cdot \opt(I)+1$, where $\lambda\in(0,1)$.
\label{lem:eta3:robustness:approx}
\end{lemma}

\begin{proof}
We start from an empty set $Q$.
Suppose that the last time point in the current $Q$ is $q_{i}$.
In each time point $t\geq q_{i}+1$, we can compute the value of $\opt(I\ang{q_{i+1},t})$.
As long as the time point $t$ satisfies (\rom{1}) $\opt(I\ang{q_{i+1},t})\leq\frac{1}{\lambda}$; (\rom{2}) $\opt(I\ang{q_{i+1},t+1})>\frac{1}{\lambda}$, the time point $t$ will be added to the set $Q$.

Let $\cH_{Q}=(H_1,\ldots,H_k)$ be the solution partition of $Q$.
Note that $k\leq\ceil{\lambda\opt(I)}$.
Let $I_i=(p_t)_{t\in H_i}$.
By \cref{claim:vertical}, we have $\sum_{i=1}^{k}\opt(I_i)\leq\opt(I)+k\leq(1+\lambda)\opt(I)+1$.
\end{proof}
\begin{algorithm}[htb]
\caption{\hspace{-2pt}{\bf .} Robustness Guarantee in One Phase}
\label{alg:eta3:robustness}
\begin{algorithmic}[1]
\REQUIRE Arbitrary instance $I=(p_t)_{t\in[n]}$ of \DTAP/, its predicted instance $\hatI=(\hat{p}_t)_{t\in[n]}$, the time point set $Q$ and a constant $\lambda$.
\ENSURE A feasible solution.
\WHILE{$i\leq k$}
\FOR{each time point $t$ in $H_i$}
\IF{$\eta_{q_{i-1}+1,t}\leq \epsilon_{\lambda}$}
\STATE {\color{gray}// Small error, trust the predictions}
\STATE Run \cref{alg:eta3:general-case} for the instance $I\ang{q_{i-1}+1,t}$.
\ELSE 
\STATE {\color{gray}// Big error, follow the online algorithm}
\STATE Run the traditional online algorithm for the remaining instance $I\ang{t,q_i}$.
\ENDIF
\ENDFOR
\STATE $i\leftarrow i+1$.
\ENDWHILE
\end{algorithmic}
\end{algorithm}

Given an arbitrary instance $I$ of \DTAP/, let $Q=\set{q_1,\ldots,q_k}$ be the time point set that satisfies the property stated in \cref{lem:eta3:robustness:approx}. 
By the proof of \cref{lem:eta3:robustness:approx}, one could easily get the set $Q$ online, i.e., without knowing the instance prior.
To simplify the description of the robustness algorithm, we assume that the set $Q$ is known by the algorithm.
Let $\eta_{t,t'}$ be the error between $I\ang{t,t'}$ and $\hatI\ang{t,t'}$.
Let $\cH_Q=\set{H_1,\cdots,H_k}$ be the solution partition of $H$.
Let $I_i=(p_t)_{t\in H_i}$.
Assume that the first time point is $t_i$ and the last time point is $t_i'$ in $H_i$.
Let $\eta_i=\eta_{t_i,t_i'}$.
We first show the robustness guaranty in each $H_i$ (\cref{lem:eta3:ratio:robustness:interval}).

\begin{lemma}
Fix an arbitrary $H_i\in\cH_Q$.
Given an arbitrary instance $I=(p_t)_{t\in[n]}$ of \DTAP/, let $\alg(I)$ be the solution returned by \cref{alg:eta3:robustness}, then we have $\alg(I_i)\leq\min\{(1+\epsilon)\cdot\opt(I_i)+O(\frac{1}{\epsilon^2})\cdot\eta_i,2\cdot\opt(I)+\epsilon\}$.
\label{lem:eta3:ratio:robustness:interval}
\end{lemma}

\begin{proof}
Fix an arbitrary set $H_i\in\cH_Q$.
Suppose that \cref{alg:eta3:robustness} is switched to the traditional $2$-competitive online algorithm at the time $t$.
By our robustness condition, we know that the current error is greater than $\epsilon_{\lambda}$.
Let $S_1=\set{t'\in H_i\mid t'\leq t}$ and $S_2=H_i\setminus S_1$.
Let $I_1=(p_t)_{t\in S_1}$ and $I_2=(p_t)_{t\in S_2}$.
Thus, for each $H_i$, we have $\alg(I_1)=2\cdot\opt(I_1)+1$ and $\alg(I_2)\leq 2\cdot \opt(I_2)$.
By \cref{claim:vertical}, we have $\alg(I_i)\leq 2\cdot \opt(I_i)+3$.
By \cref{thm:eta3:ratio}, we have:
\begin{align}
\alg(I_i)\leq
\begin{cases}
\frac{1+\lambda}{1-\lambda}\cdot\opt(I_i)+\frac{1}{\epsilon_{\lambda}}\eta_i & \text{ if $\eta_i\leq\epsilon_{\lambda}$} \\
2\opt(I_i)+3 & \text{ if $\eta_i>\epsilon_{\lambda}$}
\end{cases}
\label{equ:eta3:robustness:org}
\end{align}
By \cref{lem:eta3:robustness:approx}, we have
$\opt(I_i)\leq\frac{1}{\lambda}$.
Then, we have  $\frac{1}{\epsilon_{\lambda}}\leq\frac{1}{\epsilon_{\lambda}}\cdot\left( (2-\frac{1+\lambda}{1-\lambda})\opt(I_i)+3 \right)\leq O(\frac{1}{\lambda\cdot\epsilon_{\lambda}})$.
Thus, to make the piecewise function above continuous, we can increase the factor $O(\frac{1}{\epsilon})$ to $O(\frac{1}{\epsilon^2})$.
Therefore, \cref{equ:eta3:robustness:org} can be  written as:
$$
\alg(I_i)\leq\min \{(1+\epsilon)\cdot\opt(I_i)+O(\frac{1}{\epsilon^2})\cdot\eta_i,2\opt(I_i)+3\}
$$
\end{proof}

\begin{proof}[Proof of \cref{lem:eta3:ratio:robustness}]
By \cref{lem:eta3:robustness:approx}, we have $\sum_{i}\opt(I_i)\leq(1+\lambda)\cdot\opt(I)+1$.
Thus, we have $\alg(I)\leq(1+\epsilon)\cdot\opt(I)+O(\frac{1}{\epsilon^2})\cdot\eta$.
By \cref{lem:eta3:robustness:approx} and \cref{claim:vertical}, we have $\alg(I)\leq 2(1+\lambda)\opt(I)+3\lambda\opt(I)=(2+5\lambda)\cdot\opt(I)$.
\end{proof}

\section{Learnability}
    \label{sec:learnable}

This section shows that we can learn a prediction with approximately minimum expected error given only a polynomial number of samples.
We make standard assumptions that the number of time steps is at most $T$, the number of packages per time step is at most $K$, and each DAP instance is sampled from an unknown distribution $\cD$.
Use $\cI$ to denote the set of all potential predictions $\hatI$, i.e., $\cI:=\set{(p_1,p_2,\ldots,p_T)\mid\forall t\in [T], p_t\in [0,K]}$. 

\begin{theorem}\label{thm:learnability}
For any $\epsilon,\delta \in (0,1)$ and any distribution $\cD$, after observing $O((\frac{T}{\epsilon})^2(T\log(\frac{KT}{\epsilon d})+\ln(\frac{1}{\delta}))$ samples, there exists a learning algorithm that returns a predicted instance $\hatI \in \cI$ such that with probability at least $1-\delta$,
    $\E_{I \thicksim \cD}[\eta(\hatI,I)] \leq \E_{I \thicksim \cD}[\eta(I^*,I)] + \epsilon,$
    where for any two instances $I_1,I_2$, $\eta(I_1,I_2)$ represents the error when $I_1$ is the prediction of $I_2$, and $I^* = \argmin_{I'\in \cI}\E_{I \thicksim \cD}[\eta(I',I)]$.
\end{theorem}

The basic proof idea is that we first show that with only a small loss on the minimum expected error, $\cI$ can be reduced to a finite set.
Then due to the bounded pseudo-dimension of that finite set, the sample complexity can be bounded.
Before stating the reduction, we prove a simple lemma first. 

\begin{lemma}\label{lem:l1_norm}
For any two instances $I_1=\{p_t^{(1)}\}_{t\in[T]}$ and $I_2=\{p_t^{(2)}\}_{t\in[T]}$, $|\opt(I_1)-\opt(I_2)|\leq \rho T/d$, where $\rho$ is the $\ell_1$-norm distance between the two instances. 
\end{lemma}
\begin{proof}
Let $X^{(1)}$ be the optimal solution of instance $I_1$. 
Let $\num X^{(1)}$ be the number of acknowledgements in solution $X^{(1)}$, and for each time point $t\in [T]$.
Let $D_t^{(1)}$ denote the waiting time of the packages arriving at time $t$ in solution $X^{(1)}$. We have
\begin{equation*}
    \begin{aligned}
    \opt(I_2) &\leq \sF(X^{(1)},I_2)\\
    &=\num X^{(1)} + \sum_{t\in [T]} P_t^{(2)}D_t^{(1)}/d\\
    &\leq \num X^{(1)} + \sum_{t\in [T]} P_t^{(1)}D_t^{(1)}/d + \rho T/d\\
    &=\opt(I_1) + \rho T/d.
    \end{aligned}
\end{equation*}

Analogously, we can also prove that $\opt(I_1)\leq \opt(I_2)+\rho T/d$.

\end{proof}

\begin{lemma}\label{lem:reduce_finite_set}
For any $\epsilon\in (0,1)$ and any distribution $\cD$, there exists a finite set $\cF$ with size $O((\frac{KT^2}{\epsilon d})^T)$ such that
\[\min_{\hatI \in \cF}\E_{I \thicksim \cD}[\eta(\hatI,I)] \leq \min_{\hatI \in \cI}\E_{I \thicksim \cD}[\eta(\hatI,I)] + \epsilon. \]
\end{lemma}

\begin{proof}
Given an $\epsilon\in (0,1)$, define $\epsilon_0:=\epsilon d/T^2$.
The finite set $\cF$ is defined to be a set of instance $I \in \cI$, in which the number of packages at any time point is an integral multiple of $\epsilon_0$. Clearly, the size of $\cF$ is $O( (K/\epsilon_0)^T)$. Now we show that for any instance $I\thicksim \cD$ and $\hatI\in \cI/\cF$, we can round instance $\hatI$ to an instance $\hatI'\in \cF$ such that $|\eta(\hatI,I)-\eta(\hatI',I)|\leq \epsilon$, which implies $\min_{\hatI \in \cF}\E_{I \thicksim \cD}[\eta(\hatI,I)] \leq \min_{\hatI \in \cI}\E_{I \thicksim \cD}[\eta(\hatI,I)] + \epsilon$ directly.

Partition the interval $[0,K]$ into $\lceil K/\epsilon_0 \rceil$ sub-intervals $[0,\epsilon_0],(\epsilon_0,2\epsilon_0],...,((\lceil K/\epsilon_0 \rceil-1)\epsilon_0, \lceil K/\epsilon_0 \rceil\epsilon_0]$. For an instance $\hatI \in \cI/\cF$, we round each $p_t$ to the upper bound of the sub-interval that $p_t$ belongs to. Denote the new instance after rounding by $\hatI'$. Clearly, the $\ell_1$-norm distance between instance $\hatI$ and $\hatI'$ is at most $T\epsilon_0$.

For convenience, let $I_i^O:=\dO(I_i,\hatI_i)$ and $I_i^U:=\dU(I_i,\hatI_i)$.
Recall that 
$$
\eta(\hatI, I) := \max_{P\in \cP} \sum_{I_i\in P}(\opt(I_i^O)-\opt(I_i^U)).
$$
We define $P_0$ to be the partition which yields the value of $\eta(\hatI', I)$, i.e., 
$$
P_0 =\argmax_{P\in \cP} \sum_{I_i\in P}(\opt((I_i')^O)-\opt((I_i')^U)).
$$
Note that the $\ell_1$-norm distance between instance $\hatI$ and $\hatI'$ is at most $T\epsilon_0$.
By \cref{lem:l1_norm}, we have
\begin{equation*}
    \begin{aligned}
    \eta(\hatI',I)&= \sum_{I_i \in P_0}(\opt((I_i')^O)-\opt((I_i')^U)) \\
    &\leq \sum_{I_i \in P_0}(\opt((I_i)^O)-\opt(I_i^U)) + T^2\epsilon_0/d\\
    & \leq \eta(\hatI, I) + \epsilon.
    \end{aligned}
\end{equation*}
We can also prove $\eta(\hatI,I)\leq \eta(\hatI',I)+\epsilon$ analogously. Finally, we have $|\eta(\hatI,I)-\eta(\hatI',I)|\leq \epsilon$, which completes the proof.
\end{proof}

Since $\cF$ is a finite set with size $O((\frac{KT^2}{\epsilon d})^T)$, its pseudo-dimension is $O(T\log(\frac{KT}{\epsilon d}))$. According to the theorem proposed in~\cite{DBLP:journals/siamcomp/GuptaR17}, we have the following lemma.

\begin{lemma}\label{lem:finite_set_small_dimension}
For any $\epsilon,\delta \in (0,1)$ and any distribution $\cD$, after observing $O((\frac{T}{\epsilon})^2(T\log(\frac{KT}{\epsilon d})+\ln(\frac{1}{\delta}))$ training samples, 
the predicted instance $\hatI \in \cF$ with the minimum empirical error $\eta$ among the training samples satisfies
    \[\E_{I \thicksim \cD}[\eta(\hatI,I)] \leq \E_{I \thicksim \cD}[\eta(I^*,I)] + \epsilon \]
    with probability at least $1-\delta$, where $I^* = \argmin_{I'\in \cF}\E_{I \thicksim \cD}[\eta(I',I)]$.
\end{lemma}

Combining \cref{lem:reduce_finite_set} and \cref{lem:finite_set_small_dimension}, \cref{thm:learnability} can be proved directly.
\section{Experiments}
\label{sec:experiment}

This section empirically validates our adaptive learning-augmented algorithm (ALA) 's efficiency. We investigate various types of input distribution and show superior performance compared to previous algorithms.

\subsection{Setup}

\paragraph{Input distributions and Noisy Predictions}
The experiments follow the setting in~\cite{DBLP:conf/nips/BamasMS20}. We set the delay factor $d=100$ and the maximum number of time steps $T=1000$. For each time point, the number of demands is i.i.d. sampled from a given distribution $\cD$. 
We investigate the same distributions as in~\cite{DBLP:conf/nips/BamasMS20}: the Poisson distribution of mean~$1$, the Pareto distribution of shape~$2$ and the iterated Poisson distribution of mean~$1$.
The iterated Poisson distribution is a custom distribution introduced in~\cite{DBLP:conf/nips/BamasMS20}, which is iterating on sampling a value from the Poisson distribution whose mean is the sampled value in the last iteration (initially, the mean is $1$).
The prediction of an instance is constructed by perturbing it with some noise. For each time $t$, there are two operations: setting $p_t = 0$ and adding to $p_t$ a random noise sampled from $\cD$. 
We perform each operation sequentially and independently with probability $r\in [0,1]$ at each time point.
Then, the perturbed instance is served as the prediction. 
The perturbing probability can be viewed as a simplified measure of the prediction error. The experiments test the performance of algorithms over different perturbing probabilities under each input distribution.

\paragraph{Baseline Algorithms} In addition to our algorithm, which is parameterized by $\lambda$, we implemented the following algorithms for comparison:

\begin{itemize}
    \item \greedy~\cite{DBLP:conf/stoc/DoolyGS98}. This algorithm acks when the cumulative delay cost equals 1 (the ack cost). It is the best deterministic algorithm without predictions.
    \item \pd~\cite{DBLP:conf/nips/BamasMS20}. This algorithm is the first algorithm incorporating predictions. The algorithm has a control parameter $\beta \in (0, 1]$, and has a better consistency guarantee with a smaller value of $\beta$ at the cost of a worse robustness guarantee. This was termed as the primal dual learning-augmented algorithm by the authors. 
    \item \blind. This algorithm follows the prediction blindly. It applies the optimal solution on the predicted instance to the actual instance with no adaptation.
    
\end{itemize}

Note that we did not include the $\frac{e}{e-1}$-competitive randomized algorithm as it rarely outperforms the  2-competitive deterministic algorithm in practice.


\paragraph{Computational Settings.} We conduct experiments on a machine running Ubuntu 18.04 with an i7-7800X CPU and 48 GB memory. The algorithms are implemented in Python 3.8, and the results are averaged over five runs. 

\subsection{Empirical Discussion}

\begin{figure*}[tb]
\centering
\subfigure[$\beta=1,\lambda=0.58$]{
    \centering
    \includegraphics[width=0.3\textwidth]{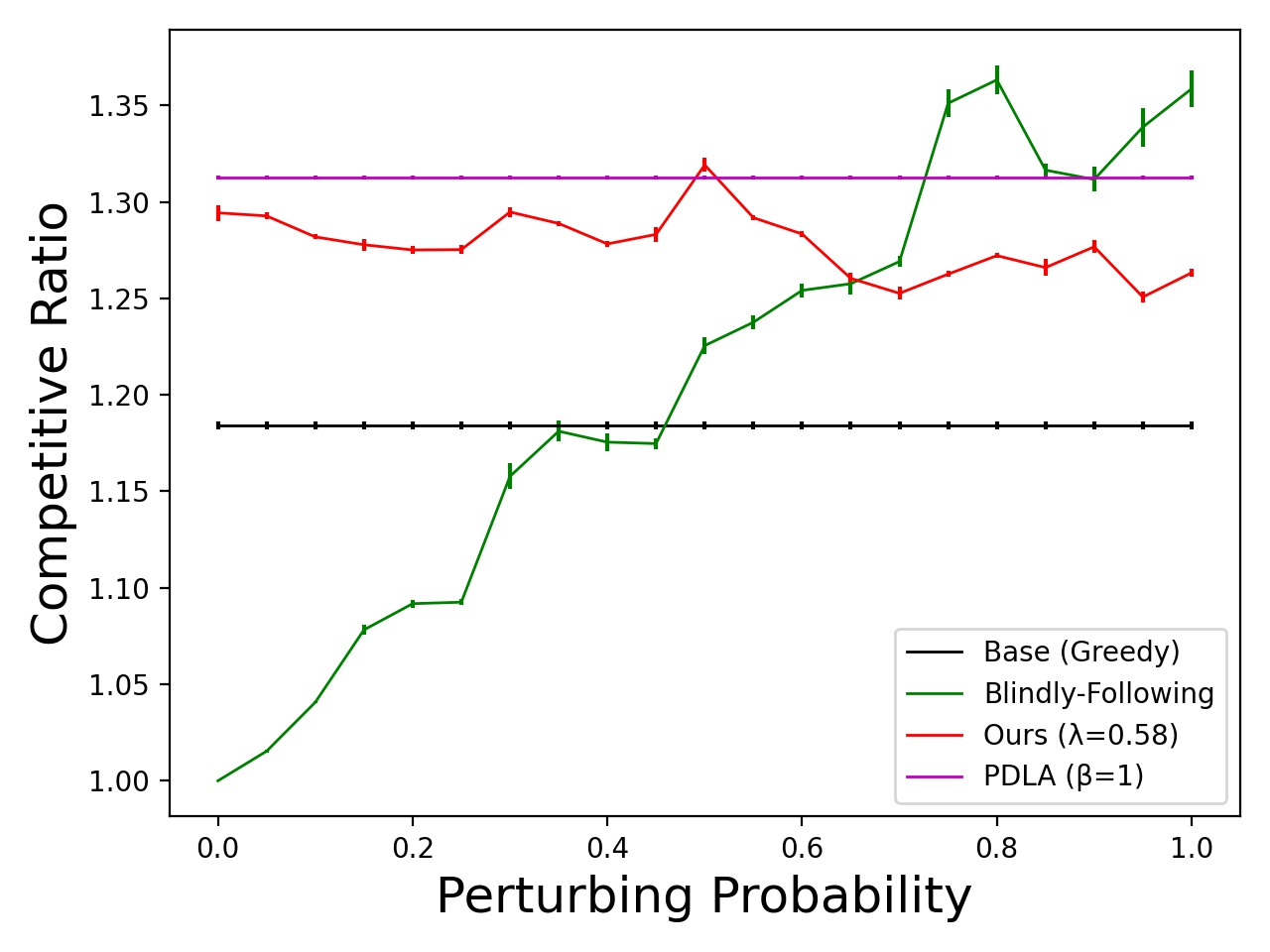}
    \label{fig:Pareto_0}
    }
\subfigure[$\beta=0.6,\lambda=0.32$]{
    \centering
    \includegraphics[width=0.3\textwidth]{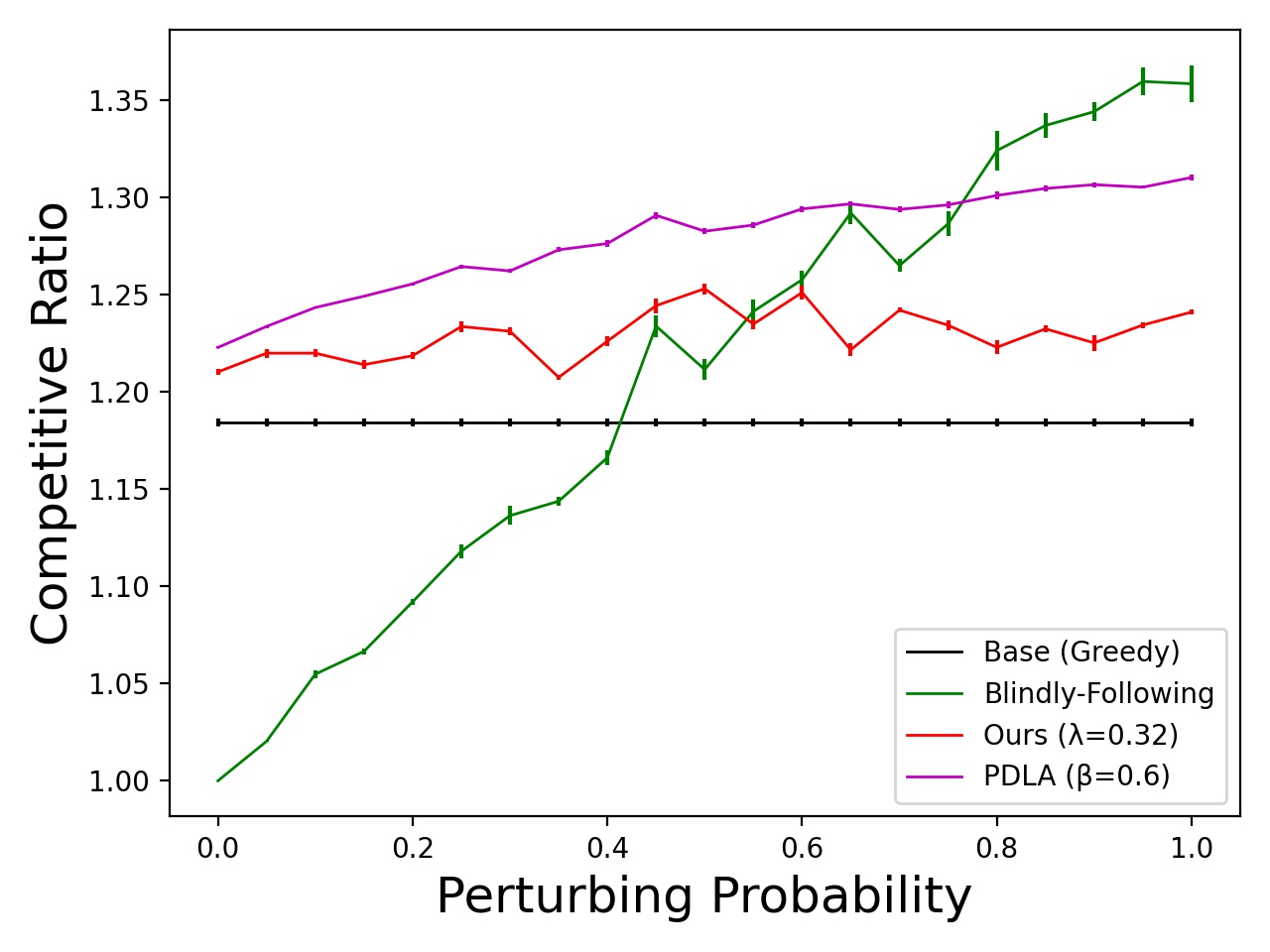}
    \label{fig:Pareto_1}
    }
\subfigure[$\beta=0.2, \lambda=0.1$]{  
    \centering
    \includegraphics[width=0.3\textwidth]{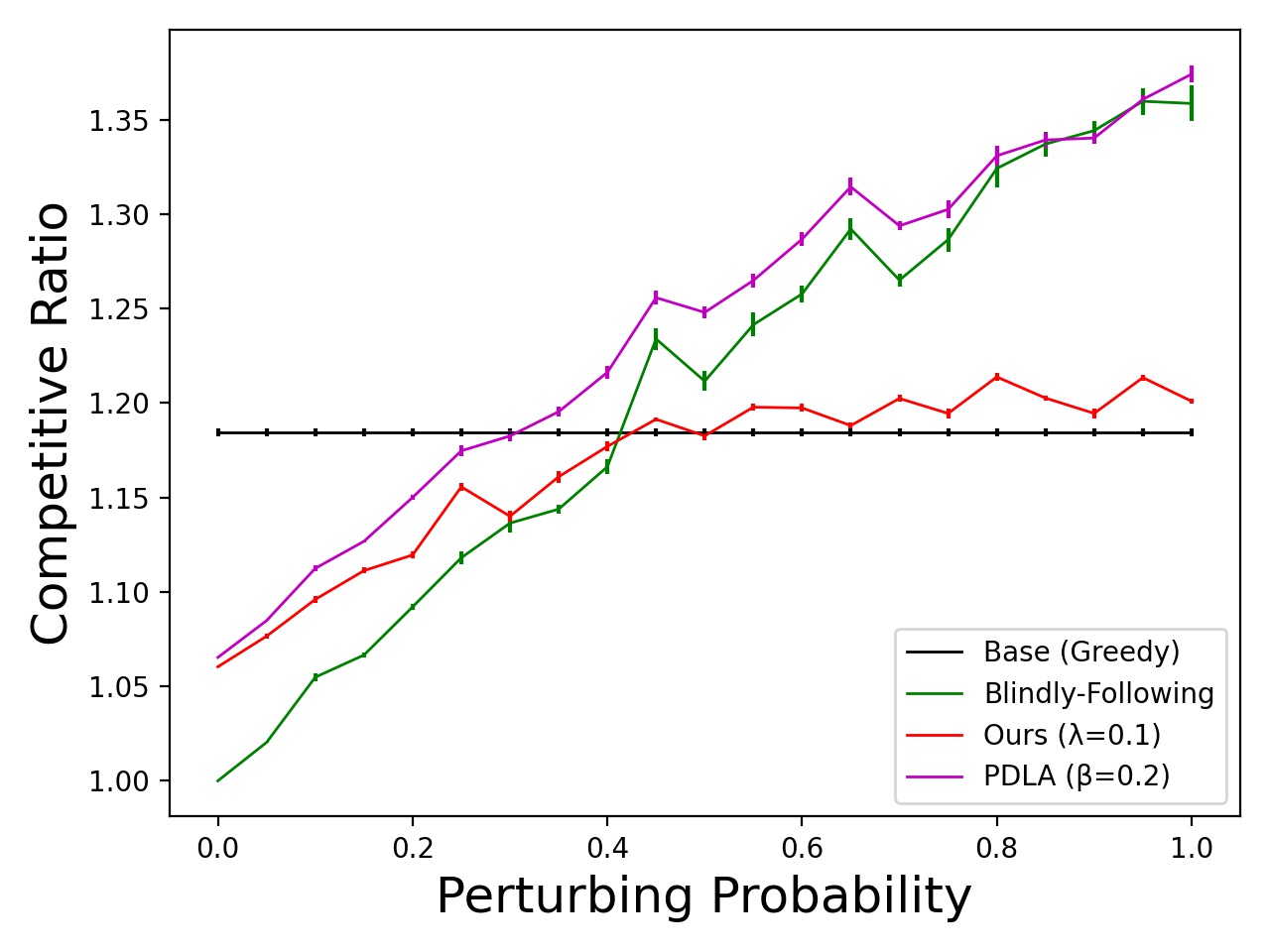}
    \label{fig:Pareto_2}
    }
    \caption{The performance of the algorithms under the Pareto distribution. A large perturbing probability implies a large prediction error. Recollect that $\pd$~\cite{DBLP:conf/nips/BamasMS20} parameterized by $\beta$ has a consistency ratio of $\frac{\beta}{1-e^{-\beta}}$, and our algorithm parameterized by $\lambda$ a consistency ratio of $(1+\lambda)$. For fair comparison, we consider different pairs of $(\beta,\lambda)$ that give the same consistency ratio for $\pd$ and our algorithm. 
    }
    \label{fig:exp}
\end{figure*}

\begin{figure*}[htbp]
\centering
\subfigure[$\beta=1,\lambda=0.58$]{
    \centering
    \includegraphics[width=0.3\textwidth]{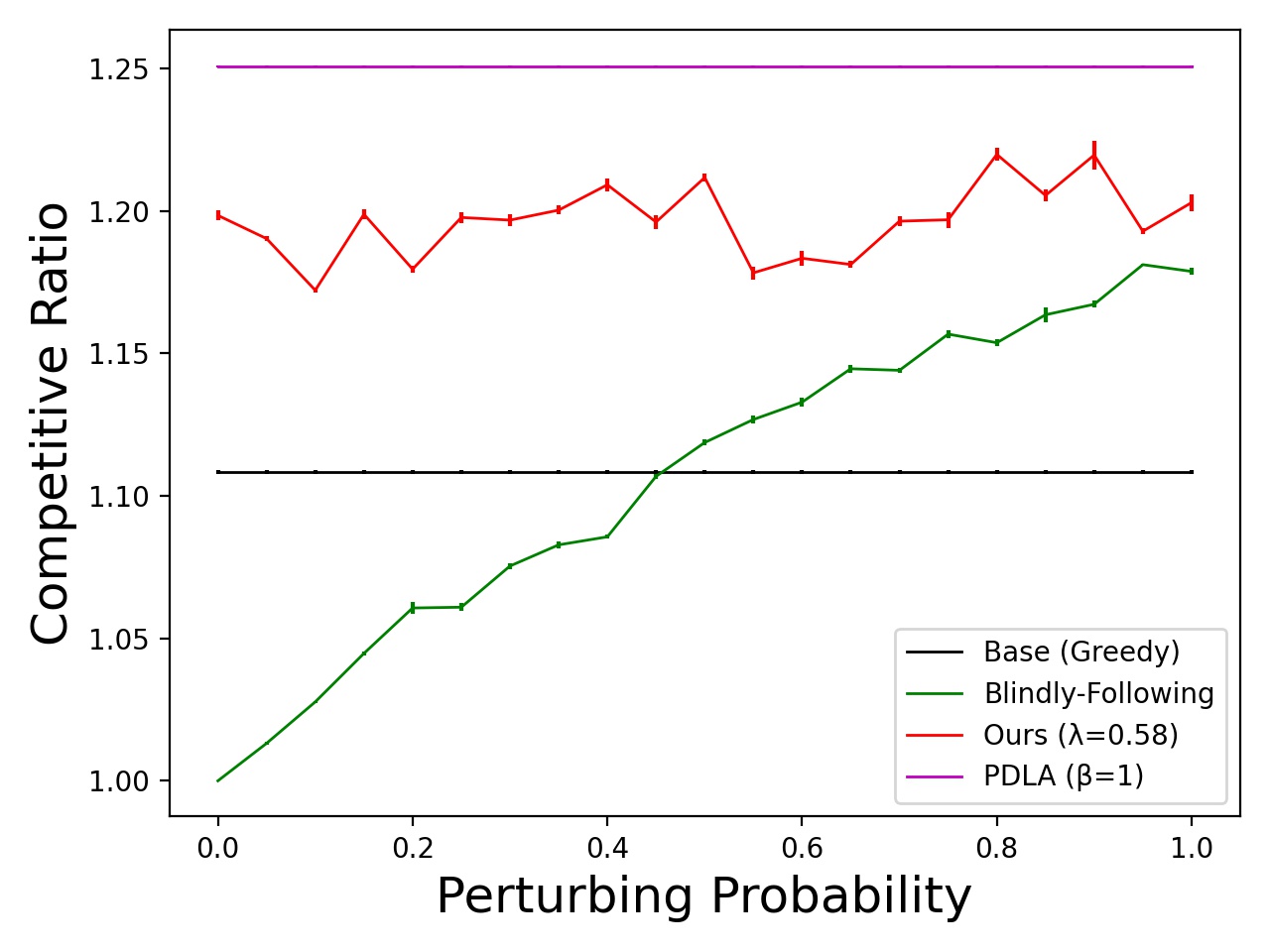}
    \label{fig:Poisson_0}
    }
\subfigure[$\beta=0.6,\lambda=0.32$]{
    \centering
    \includegraphics[width=0.3\textwidth]{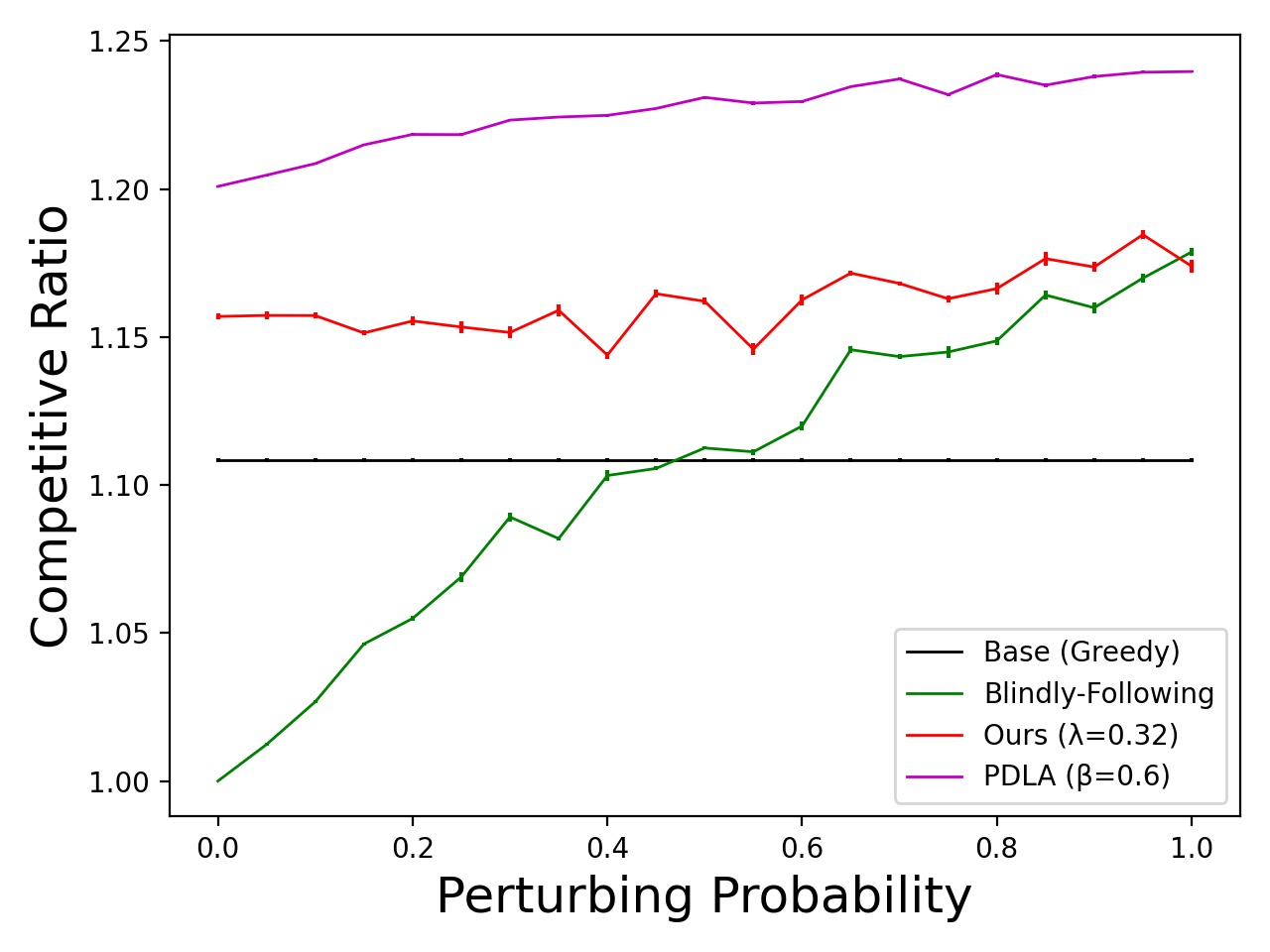}
    \label{fig:Poisson_1}
    }  
\subfigure[$\beta=0.2, \lambda=0.1$]{  
    \centering
    \includegraphics[width=0.3\textwidth]{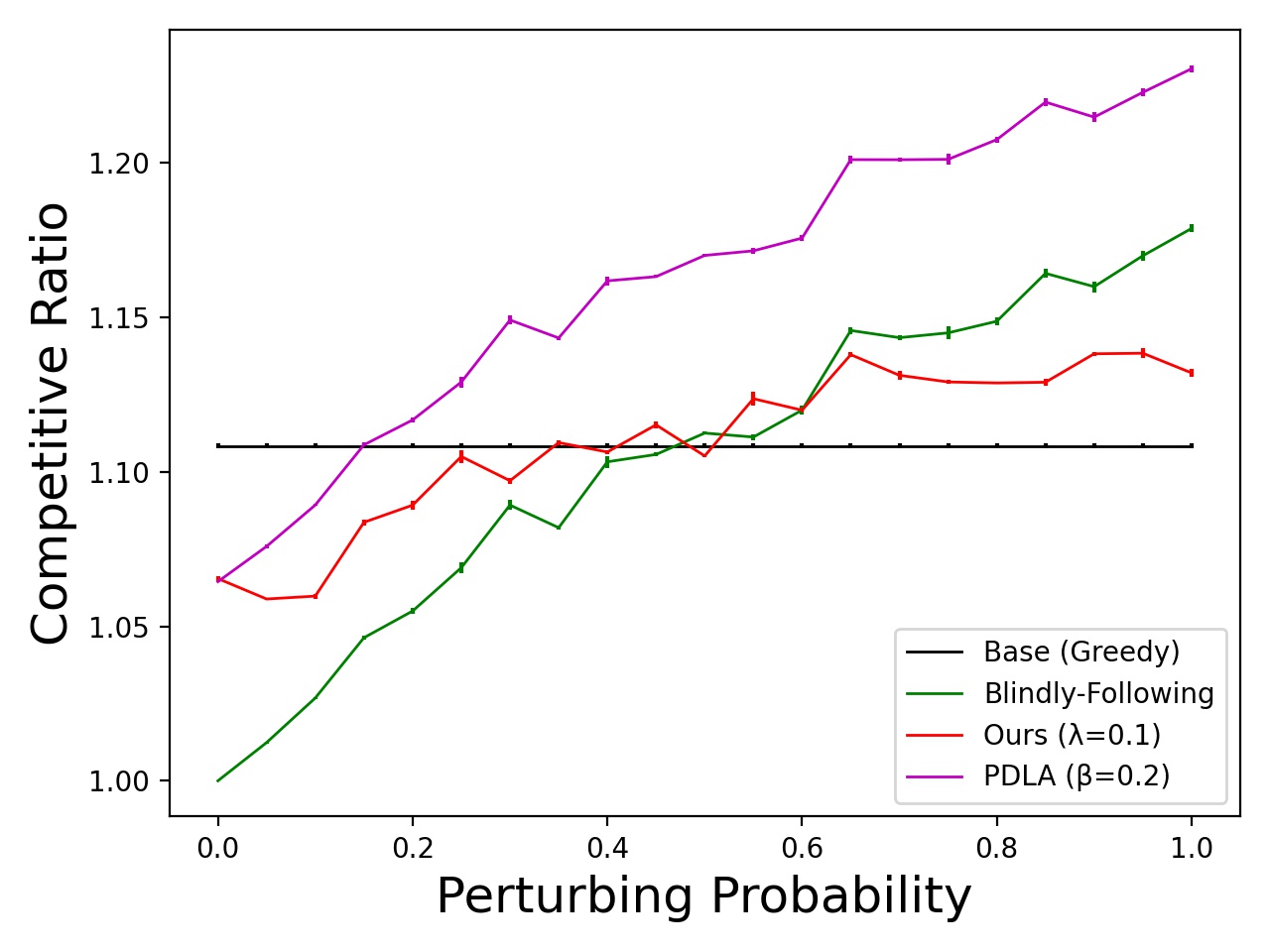}
    \label{fig:Poisson_2}
    }
    \caption{The performance of algorithms under the Poisson distribution.
    }
    \label{fig:Poisson}
\end{figure*}

\begin{figure*}[tb]
\centering
\subfigure[$\beta=1,\lambda=0.58$]{
    \centering
    \includegraphics[width=0.3\textwidth]{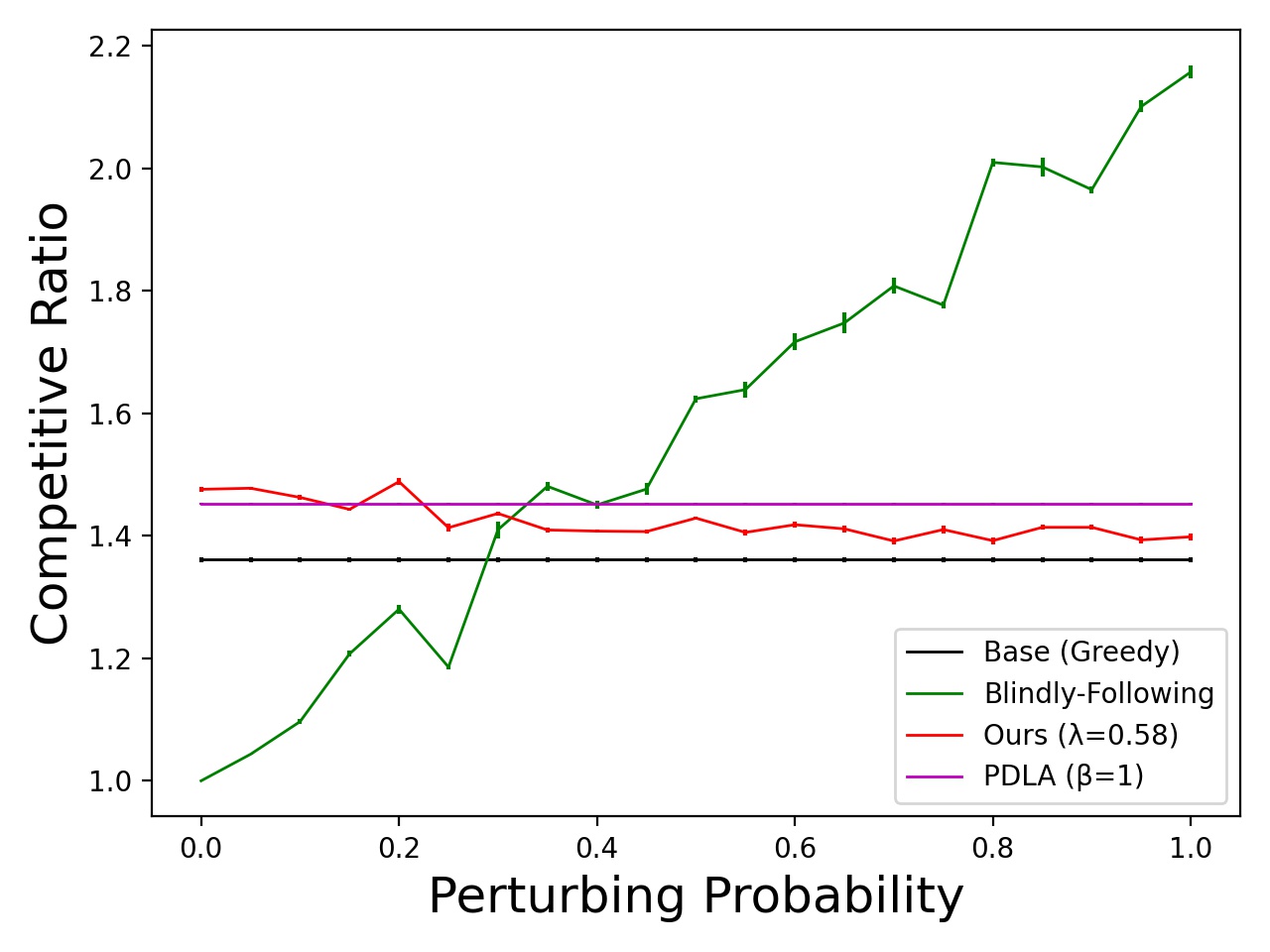}
    \label{fig:Iterated_Poisson_0}
    }
\subfigure[$\beta=0.6,\lambda=0.32$]{
    \centering
    \includegraphics[width=0.3\textwidth]{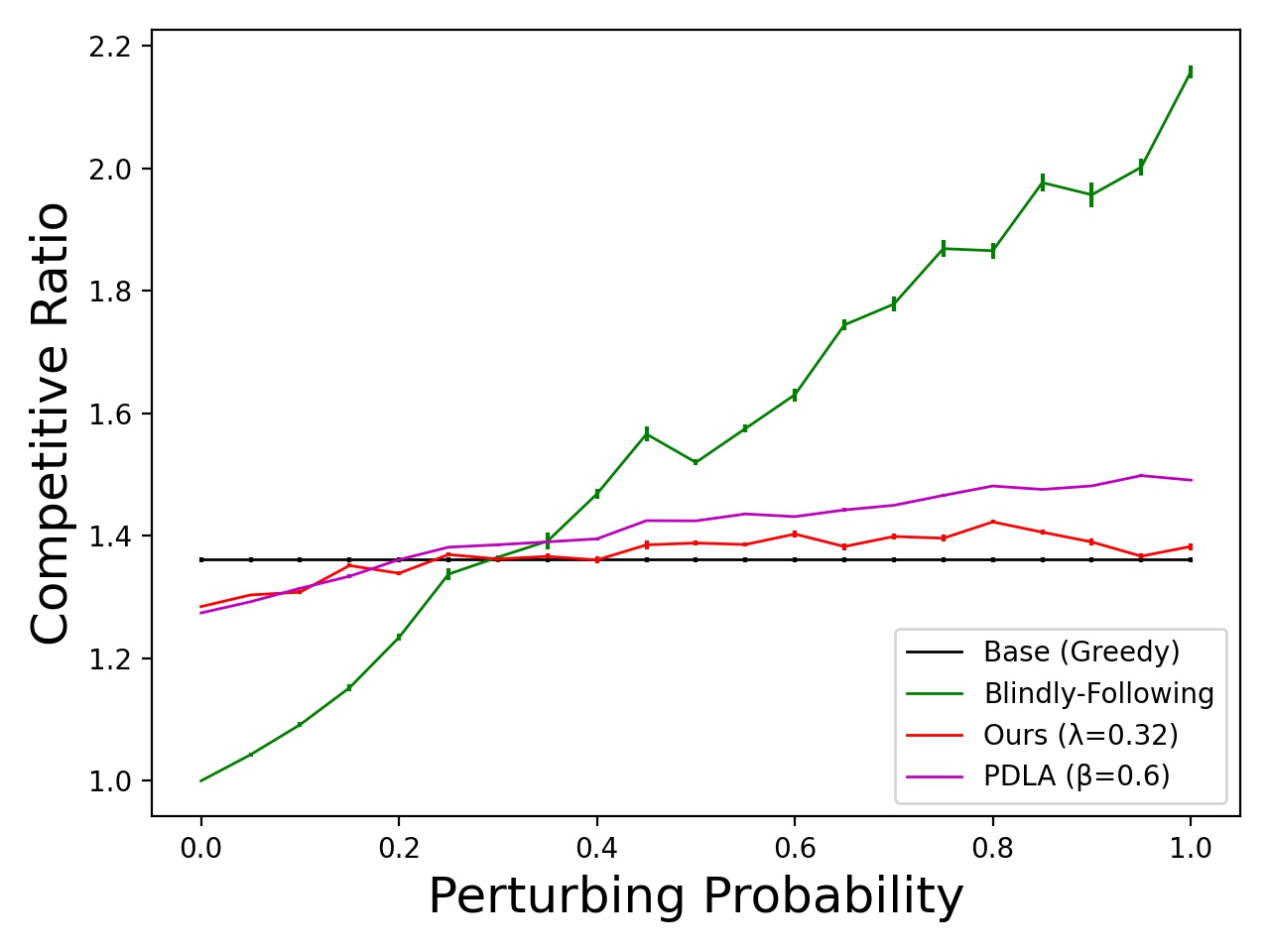}
    \label{fig:Iterated_Poisson_1}
    }
\subfigure[$\beta=0.2, \lambda=0.1$]{  
    \centering
    \includegraphics[width=0.3\textwidth]{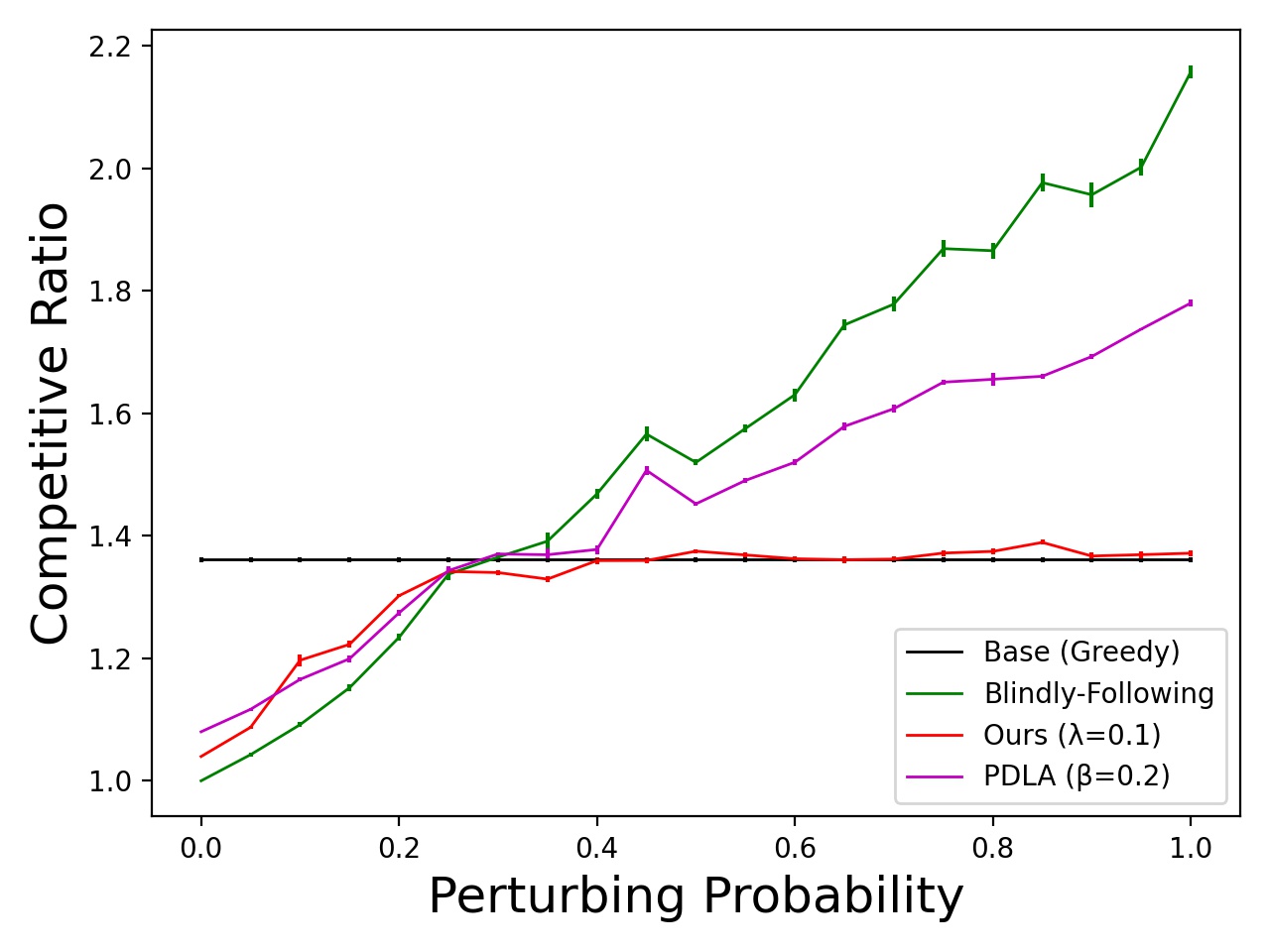}
    \label{fig:Iterated_Poisson_2}
    }
    \caption{The performance of algorithms under the iterated Poisson distribution.
    }
    \label{fig:exp_iterated_possion}
\end{figure*}

All the results for the three considered distributions exhibit similar patterns. 
Thus, we only discuss the results for the Pareto distribution in \cref{fig:exp}, and only show the results for the Poisson distribution (\cref{fig:Poisson}) and the iterated Poisson distribution (\cref{fig:exp_iterated_possion}). 
We observe the following.

\begin{itemize}
    \item Our algorithm is robust to the error. Further, with a small $\lambda$, our algorithm has a better empirical competitive ratio than algorithm $\greedy$ when the prediction error (perturbing probability) is small and remains on par with it, regardless of the error. 
    
    \item For the choice of $\beta$ and $\lambda$ values that lead to the same consistency guarantee, our algorithm shows a better performance than $\pd$ in most cases. 
    This demonstrates that our algorithm obtains a better trade-off between consistency and robustness, confirming the importance of simultaneous optimum consistency and robustness achieved by our algorithm.
    
\end{itemize}

\section{Conclusion}

This paper revisited the dynamic acknowledgment problem. For this problem, previously, it was unclear what a good error measure should be in the learning augmented algorithm analysis model. One of this paper's main contributions lies in formulating a novel error measure and designing algorithms based on the error. The algorithm developed in this paper achieves simultaneous optimum consistency and robustness, the most desirable result. The theory is verified empirically. We believe our new error and algorithm could inspire new ML-augmented solutions for other problems with a temporal nature to their input\footnote{The authors have provided public access to their code at \url{https://github.com/Chenyang-1995/TCP}}.

\section*{Acknowledgement}
Chenyang Xu was supported in part by Science and Technology Innovation 2030 –``The Next Generation of Artificial Intelligence" Major Project No.2018AAA0100900.
Sungjin Im was supported in part by NSF grants CCF-1844939 and CCF-2121745. Benjamin Moseley was supported in part by  a Google Research Award, an Infor Research Award, a Carnegie Bosch Junior Faculty Chair, and NSF grants CCF-2121744 and  CCF-1845146

\newpage
\clearpage
\printbibliography

\newpage

\appendix

\end{document}